\documentclass[]{pasj02} 
\usepackage{booktabs}
\usepackage{tabularx}
\usepackage {diagbox}
\usepackage{multirow}
\usepackage{xcolor}
\usepackage{url}
\usepackage{lmodern}
\usepackage[T1]{fontenc}  
\usepackage{siunitx}
\usepackage{mleftright}
\usepackage{blindtext}
\usepackage{bm}

\jyear{2024}
\Received{}
\Accepted{}

\begin{document}

\title{Infrared Bubble Recognition in the Milky Way and Beyond Using Deep Learning }
\author{Shimpei \textsc{Nishimoto},\altaffilmark{1}\footnotemark[*]
Toshikazu \textsc{Onishi},\altaffilmark{1}
Atsushi \textsc{Nishimura},\altaffilmark{2}
Shinji \textsc{Fujita},\altaffilmark{3}
Yasutomo \textsc{Kawanishi},\altaffilmark{4}
Shuyo \textsc{Nakatani},\altaffilmark{5}
Kazuki \textsc{Tokuda},\altaffilmark{6,7}
Yoshito \textsc{Shimajiri},\altaffilmark{8}
Hiroyuki \textsc{Kaneko},\altaffilmark{9}
Yusuke \textsc{Miyamoto},\altaffilmark{10}
Tsuyoshi \textsc{Inoue},\altaffilmark{11} and 
Atsushi \textsc{M. Ito},\altaffilmark{12}
}

\altaffiltext{1}{Department of Physics, Graduate School of Science, Osaka Metropolitan University, 1-1 Gakuen-cho, Naka-ku, Sakai, Osaka 599-8531, Japan}
\altaffiltext{2}{National Astronomical Observatory of Japan (NAOJ), Nobeyama Radio Observatory  462-2 Minamimaki-mura, Saku, Nagano , 384-1305, Japan}
\altaffiltext{3}{The Institute of Statistical Mathematics 10-3 Midori-cho,Tachikawa Tokyo 190-8562, Japan}
\altaffiltext{4}{RIKEN Information R\&D and Strategy Headquarters, 2-2-2 Hikaridai, Seika-cho, Soraku-gun, Kyoto, 619-0288, Japan}
\altaffiltext{5}{Cybozu Labs Inc, Tokyo Nihombashi Tower 27F, 2-7-1 Nihombashi, Chuo-ku, Tokyo 103-6027, Japan}
\altaffiltext{6}{Kyushu University, 6-10-1 hakozaki, higashi-ku, Fukuoka City, 812-8581, JAPAN}
\altaffiltext{7}{National Astronomical Observatory of Japan, National Institutes of Natural Sciences, 2-21-1 Osawa, Mitaka, Tokyo 181-8588, Japan}
\altaffiltext{8}{Kyushu Kyoritsu University, 1-8 Jiyugaoka, Yahatanishi-ku, Kitakyushu-shi, Fukuoka, 807-8585, Japan}
\altaffiltext{9}{Department of Physical Science, Niigata University, 8050, Ikarashi 2-no-cho, Nishi-ku, Niigata 950-2181, Japan}
\altaffiltext{10}{Fukui University of Technology, Faculty of Engineering, Development of Electronic Engineering, 3-6-1, Gakuen, Fukui, Fukui, 910-8505, Japan}
\altaffiltext{11}{Department of Physics, Konan University, 8-9-1 Higashinada, Kobe, Hyogo, 658-8501, Japan}
\altaffiltext{12}{National Institute for Fusion Science (NIFS), National Institutes of Natural Sciences (NINS) 322-6 Oroshi, Toki, Gifu, 509-5292, Japan}

\email{shimpei.nishimoto.astronomy@gmail.com}

\maketitle

\begin{abstract}

We propose a deep learning model that can detect Spitzer bubbles accurately using two-wavelength near-infrared data acquired by the Spitzer Space Telescope and JWST. The model is based on the Single Shot MultiBox Detector as an object detection model, trained and validated using Spitzer bubbles identified by the Milky Way Project (MWP-Bubble). We found that using only MWP-Bubbles with clear structures, along with normalization and data augmentation, significantly improved performance. To reduce the dataset bias, we also use the data without bubbles in the dataset selected by combining two techniques: negative sampling and clustering. The model was optimized by hyperparameter tuning using Bayesian optimization. Applying this model to a test region of the Galactic plane resulted in a 98 $\%$ detection rate for MWP-Bubbles with \SI{8}{\micro m} emission clearly encompassing \SI{24}{\micro m} emission. Additionally, we applied the model to a broader area of $1^\circ \leq |l| \leq 65^\circ$, $|b| \leq 1^\circ$, including both training and validation regions, and the model detected 3,006 bubbles, of which 1,413 were newly detected. We also attempted to detect bubbles in the high-mass star-forming region Cygnus $X$, as well as in the external galaxies Large Magellanic Cloud (LMC) and NGC 628. The model successfully detected Spitzer bubbles in these external galaxies, though it also detected Mira-type variable stars and other compact sources that can be difficult to distinguish from Spitzer bubbles. The detection process takes only a few hours, demonstrating the efficiency in detecting bubble structures. Furthermore, the method used for detecting Spitzer bubbles was applied to detect shell-like structures observable only in the \SI{8}{\micro m} emission band, leading to the detection of 469 shell-like structures in the LMC and 143 in NGC 628.
\end{abstract}

\section{Introduction}
High-mass stars significantly impact the surrounding interstellar medium (ISM) and the evolution of galaxies \citep{2014PhR...539...49K}. 
Mechanically, they dynamically disturb the ISM through stellar winds, ionizing radiation, dust heating, and the expansion of H{\sc ii} regions, etc \citep{2005ApJ...623..917H, 2008ApJ...681.1341W, 2014A&A...564A..68S}. Chemically, their supernova explosions at the end of their lifecycles enrich heavy elements \citep{2002ApJ...577..853W, 2013ARA&A..51..457N}. Therefore, a deep understanding of the mechanisms of the high-mass star formation is crucial for comprehending galaxy evolution.

The ISM is filled with ring-like and shell-like structures of various sizes, ranging from supergiant shells spanning kiloparsecs created by energetic events such as supernova explosions \citep{1999AJ....118.2797K} to smaller structures associated with protostar formation \citep{2023ApJ...945...63H, 2023ApJ...956L..16T}. Recent JWST observations have revealed that entire galaxies are densely packed with these ring and shell structures \citep{2023ApJ...944L..22B, 2023ApJ...944L..24W}. Unraveling the nature and formation mechanisms of these structures is key to understanding the lifecycle of the ISM in the context of star formation and ultimately deciphering the process of galaxy evolution.

Among these numerous rings and shells, Spitzer bubbles are particularly well-studied and have been systematically identified in large numbers, especially within the Milky Way \citep{2006ApJ...649..759C, 2007ApJ...670..428C, 2012MNRAS.424.2442S, 2019MNRAS.488.1141J}. These structures are characterized by an \SI{8}{\micro m} bright shell surrounding a central \SI{24}{\micro m} emission region (see Figure 1 in \cite{2014ApJS..214....3B}). Spitzer bubbles have typically been associated with the feedback effects of the high-mass stars, such as in the Collect $\&$ Collapse (C$\&$C) process \citep{1977ApJ...214..725E}. However, there is ongoing debate about their formation mechanisms \citep{2007A&A...472..835Z, 2015ApJ...806....7T}, as some evidence implies that they capture the direct triggers of high-mass star formation, as proposed in the Cloud-Cloud Collision (CCC: \cite{1992PASJ...44..203H, 2021PASJ...73S...1F} and the references therein). Understanding the origins of Spitzer bubbles is, therefore, critical for constraining the mechanisms of high-mass star formation.

In addition, Spitzer bubbles are used not only to understand individual star-forming regions, but also as a statistical study to understand the mechanisms of high-mass star formation such as CCC and C$\&$C in the entire Milky Way \citep{2012ApJ...755...71K, 2012MNRAS.421..408T}. Statistical studies of such high-mass star formation mechanisms require comprehensive and accurate detection of Spitzer bubbles in the entire Milky Way and other galaxies.
However, some Spitzer bubbles were not detected in previous studies and the conventional detection of Spitzer bubbles has primarily relied on manual work, which is time-consuming and costly \citep{2020SPIE11452E..2LU}. Therefore, we developed a deep-learning model that can detect Spitzer bubbles. Deep learning generally enables faster and more accurate detection by processing large datasets. Additionally, deep learning excels at capturing microscopic structures and patterns, helping us to better understand the physical processes involved in the formation and evolution of the Spitzer bubble, as well as its interaction with the surrounding environment. This approach also saves time and resources, making large-scale observational data analysis feasible. By comparing the spatial and velocity distribution of molecular gas associated with the Spitzer bubbles, we can statistically investigate the origins of Spitzer bubbles and the mechanisms of high-mass star formation \citep{2018PASJ...70S..51T, 2019ApJ...872...49F, 2015ApJ...798...30L}. We are preparing the following paper about investigating star formation mechanisms through comparison with molecular gas.

In this study, we focus on developing a deep learning model that can rapidly detect Spitzer bubbles including previously undetected ones. In section~\ref{sec:Data_description}, we introduce the infrared data used in this study. Section~\ref{sec:Method_of_Spitzer_bubble_detection} describes details of the model, dataset, method of data processing, and evaluation metrics. To enhance the model performance, we conducted data optimization in section~\ref{sec:Details_of_data_optimization} and model's hyperparameter tuning in section~\ref{sec:Hyperparameter_optimization_our_model}. In section~\ref{sec:Result_Discussion}, we discuss the effectiveness of the model in detecting Spitzer bubble within the test region, Cygnus $X$, the Large Magellanic Cloud (LMC), and NGC 628. We also discuss results of detecting shell-like structures observable only in the \SI{8}{\micro m}, which are considered to be generated by supernova explosions or high-mass star formations, utilizing the methods described in sections~\ref{sec:Details_of_data_optimization} and \ref{sec:Hyperparameter_optimization_our_model}.

\subsection{Feature of Spitzer bubbles}
\label{sec:Feature_of_Spitzer_bubble}
The \SI{24}{\micro m} emission associated with the Spitzer bubbles is considered to trace H{\sc ii} regions, which are ionized and heated to approximately 10$^4$ K by ultraviolet (UV) radiation from high-mass stars. Within the H{\sc ii} region, dust mixed with ionized gas is heated by an intense radiation field, forming a bright nebula at \SI{24}{\micro m} similar to radio continuum emission (\cite{2006ApJ...649..759C, 2008ApJ...681.1341W, 2010ApJ...716.1478W}). The photodissociation region (PDR) surrounding the H{\sc ii} region is traced by \SI{8}{\micro m} emission, which is produced by the excitation of polycyclic aromatic hydrocarbons (PAHs) due to far-UV radiation leaking from the H{\sc ii} region. Thus, the characteristic morphology of Spitzer bubbles, where a bright \SI{8}{\micro m} shell surrounds the central part of the \SI{24}{\micro m} emission, can be observed in both CCC and C$\&$C scenarios (\cite{2014ApJ...792...63T, 2016IAUS..315E..72S, 2007MNRAS.375.1291D}).

H{\sc ii} regions are the most abundant energy sources of turbulence within giant molecular clouds (GMCs) \citep{2002ApJ...566..302M}. Among the 102 bubbles picked up from Spitzer bubbles detected by Churchwell et al. (\yearcite{2006ApJ...649..759C},~\yearcite{2007ApJ...670..428C}) (hereafter CH06, 07), at least 86\% coincide with radio continuum emission at \SI{20}{cm} \citep{2010A&A...523A...6D}. This result suggests that most Spitzer bubbles are associated with H{\sc ii} regions and are significant energy sources within GMCs. Additionally, Spitzer bubbles are helpful as tracers of star formation activity due to their relatively low contamination from supernova remnants (SNR), asymptotic giant branch star bubbles, and planetary nebulae \citep{2010A&A...523A...6D}. Thus, Spitzer bubbles are appropriate objects as indicators of high-mass star formation and have important features for understanding the process of high-mass star formation.

\subsection{Previous work on the detection of Spitzer bubbles}
    \begin{table*}[t]
    \caption
        { \label{table:GLIMPSE_MIPSGAL} 
    Comparison of wavelengths, angular resolutions, and observation areas of GLIMPSE and MIPSGAL. In this study, we use only GLIMPSE \SI{8}{\micro m} and MIPSGAL \SI{24}{\micro m}.}
    \centering
    \begin{tabularx}{\textwidth}{@{}lXX@{}}
    \toprule
         & \textbf{GLIMPSE}  & \textbf{MIPSGAL} \\ \midrule \midrule
    \textbf{Wavelengths}   & 3.6, 4.5, 5.8, \SI{8}{\micro m}  & 24, \SI{70}{\micro m} \\
    \textbf{Resolutions}  & \timeform{1".5} -- \timeform{1".9} & 5$''$, 15$''$ \\
    \textbf{Area} & $-65^\circ \leq l \leq 65^\circ$, $-1^\circ \leq b \leq 1^\circ$ & $-65^\circ \leq l \leq 65^\circ$, $-1^\circ \leq b \leq 1^\circ$ \\ \bottomrule
    \end{tabularx}
    \end{table*}

\subsubsection{Manual work}
The mid-infrared image surveys of the Galactic plane conducted by ISO (Infrared Space Observatory; \cite{1996A&A...315L..27K}) and MSX (Midcourse Space Experiment; \cite{1995SSRv...74...81P, 1998ApJ...494L.199E, 2001AJ....121.2819P}) revealed the presence of many Spitzer bubble-like structures in the Galactic disk. Subsequently, CH06, 07 created the first Spitzer bubble catalogs, which include 591 Spitzer bubbles within the range $-65^{\circ} \leq l \leq 65^{\circ}$, $-1^{\circ} \leq b \leq 1^{\circ}$, using only the GLIMPSE data at 3.6, 5.8, and \SI{8.0}{\micro m}. However, due to the limited resources, it was suggested that the actual number of Spitzer bubbles within the surveyed area of the Galactic plane might be underestimated. Additionally, the CH06, 07 catalogs contain errors where at least two or more instances of the same Spitzer bubble were counted multiple times (for example, N1 and CN146, S1 and CS116). 

Following the above studies, the Milky Way Project (MWP), involving over 35,000 citizen scientists, detected Spitzer bubbles in the same region as CH06, 07 using GLIMPSE \SI{8}{\micro m} and MIPSGAL \SI{24}{\micro m} data (\cite{2012MNRAS.424.2442S}, DR1). Because the Spitzer bubble is associated with a H{\sc ii} region, as mentioned above, the addition of \SI{24}{\micro m} data makes the identification of Spitzer bubbles much easier than when relying primarily on \SI{8}{\micro m} data. Consequently, the project detected 5,106 Spitzer bubbles, including 86 $\%$ of the sources in CH06, 07. In addition, 928 yellow balls, which have compact emissions both in \SI{8}{\micro m} and \SI{24}{\micro m} and are expected in the early stages of high-mass star formation, were also detected \citep{2015ApJ...799..153K}. The significant increase in the number of Spitzer bubbles has enabled statistical studies of high-mass star formation on the Galactic scale. However, the DR1 is not recommended because of low accuracy in measuring the shapes and sizes of bubbles and lack of uncertainty parameters \citep{2019MNRAS.488.1141J}. Therefore, 2,600 Spitzer bubbles were newly scrutinized and cataloged (\cite{2019MNRAS.488.1141J}, DR2). The DR2 is a more refined catalog by accurately measuring the shape and size of Spitzer bubbles with a maximum zoom level that was twice that employed in the DR1 and by eliminating the duplication that existed in the DR1 (hereafter, MWP in the following text refers to the DR2).

In MWP, many scientists participated in detecting bubbles from the Spitzer data; however, this method is time-consuming, subjective, and difficult to calibrate \citep{2014ApJS..214....3B}. Despite the significant time and human resources invested in MWP, it was confirmed that undetected Spitzer bubbles still existed \citep{2020SPIE11452E..2LU}. Furthermore, with the increasing data volume from modern telescopes like the James Webb Space Telescope (JWST), comprehensive human detection is becoming increasingly difficult. Particularly, Spitzer bubble surveys by humans using all-sky data from Wide-field Infrared Survey Explorer (WISE; \cite{2010AJ....140.1868W}) or data from JWST are likely to take years. Therefore, it is crucial to detect Spitzer bubbles using machine learning with less human intervention.

\subsubsection{Machine Learning work}
Recently, machine learning techniques have been applied to various astronomical data for various scientific purposes, such as solving the Near-Far problem in the inner Galaxy using convolutional neural network (CNN) \citep{2023PASJ...75..279F} and predicting H$_2$ column density using Extra Trees Regressor which is similar to Random Forests \citep{2023MNRAS.526..966S}. 

In Spitzer bubbles, machine learning has enabled systematic, quantitative, and repeatable detection by automatically classifying them. The Random Forest classification method introduced in the DR2, named Brut, is the first introduction of the automatic classification of Spitzer bubbles (\cite{2014ApJS..214....3B, 2019MNRAS.488.1141J}). Brut has made it possible to supplement human detection by setting specific criteria for the detection of Spitzer bubbles. Subsequently, Brut significantly improved performance by training on the synthetic images of bubbles in three Spitzer bands (4.5, 8, \SI{24}{\micro m}) generated by the HYPERION (three-dimensional dust continuum Monte Carlo radiative transfer code) \citep{2017ApJ...851..149X}. However, recent advancements in object detection using CNN have shown overwhelming performance improvements compared to the Random Forest used in the Brut \citep{liu2020deep}. Focusing on CNN, \citet{2020SPIE11452E..2LU} developed a new model that can detect Spitzer bubbles (hereafter, Ueda Model), although the Ueda Model had issues such as false detections, long inference times, and complexity of result analysis. Therefore, we attempted to develop a new deep learning model that is both fast and accurate.

\section{Data description}
\label{sec:Data_description}
In this study, we used observational data from GLIMPSE, MIPSGAL, and JWST. Table~\ref{table:GLIMPSE_MIPSGAL} shows the wavelengths, resolutions, and observation areas of the GLIMPSE and MIPSGAL data.

\subsection{Galactic Legacy Infrared Mid-Plane Survey Extraordinaire (GLIMPSE)}
GLIMPSE~\citep{2003PASP..115..953B, 2009PASP..121..213C} observed the Galactic plane using the IRAC \citep{2004ApJS..154...10F} on the Spitzer Space Telescope \citep{2004ApJS..154....1W}. IRAC has four bands (3.6, 4.5, 5.8, and \SI{8.0}{\micro m}) with angular resolutions ranging from \timeform{1''.5} (\SI{3.6}{\micro m}) to \timeform{1''.9} (\SI{8.0}{\micro m}). The observational range covers $-65^{\circ} \leq l \leq 65^{\circ}$, $-1^{\circ} \leq b \leq 1^{\circ}$. In this study, we used the \SI{8.0}{\micro m} data from GLIMPSE. The \SI{8.0}{\micro m} band is dominated by strong PAH features at \SI{7.7}{\micro m} and \SI{8.6}{\micro m}, which control the diffuse emission in this band. The infrared emission from PAHs is observed in the direction of PDRs excited by far-UV radiation leaking from H{\sc ii} regions.

\subsection{MIPSGAL}
MIPSGAL is a survey of the Galactic plane ($-65^{\circ} \leq l \leq 65^{\circ}$, $-1^{\circ} \leq b \leq 1^{\circ}$) using the Multiband Imaging Photometer for Spitzer (MIPS: \cite{2004ApJS..154...25R, 2009PASP..121...76C}) at \SI{24}{\micro m} and \SI{70}{\micro m}. The angular resolutions are 5$''$ at \SI{24}{\micro m} and 15$''$ at \SI{70}{\micro m}. The \SI{24}{\micro m} emission is dominated by dust continuum emission, which is thought to be due to very small grains (VSG) out of thermal equilibrium or big grains in thermal equilibrium.

\subsection{JWST}
For the extragalactic galaxy NGC 628, we used data observed by the F770W and F2100W filters of MIRI on the JWST \citep{2006SSRv..123..485G}. The FWHM is \timeform{0''.25} for F770W and \timeform{0''.67} for F2100W. The \SI{7.7}{\micro m} band data includes PAH emission similar to GLIMPSE (\cite{2008ARA&A..46..289T}). The \SI{21}{\micro m} band data is thought to be due to thermal emission from VSG, similar to MIPSGAL. The FITS files for the data were downloaded from the Multimission Archive at STScI (MAST)\footnote{$\langle$ \url{https://mast.stsci.edu/} $\rangle$}. 

\section{Method of Spitzer bubble detection}
\label{sec:Method_of_Spitzer_bubble_detection}
To detect Spitzer bubbles at high-speed, we used one of the object detection methods, the Single Shot Multibox Detector (SSD), developed by \citet{liu2016ssd}. SSD is a CNN-based object detector that outputs the location and class confidence using a single convolutional neural network, which can speed up the object detection process. The object detection methods such as SSD are detectors that `detect' objects in images by simultaneously outputting the classification results and locations in each image. On the other hand, the Brut and the Ueda Model are classifiers that `classify' cropped images into specific categories following a sliding-window manner. Because of this feature, object detection methods are suitable for detecting in vast areas compared to classifiers.

\begin{figure*}[t]
    \begin{center}
    \begin{tabular}{c} 
    \includegraphics[width=16cm]{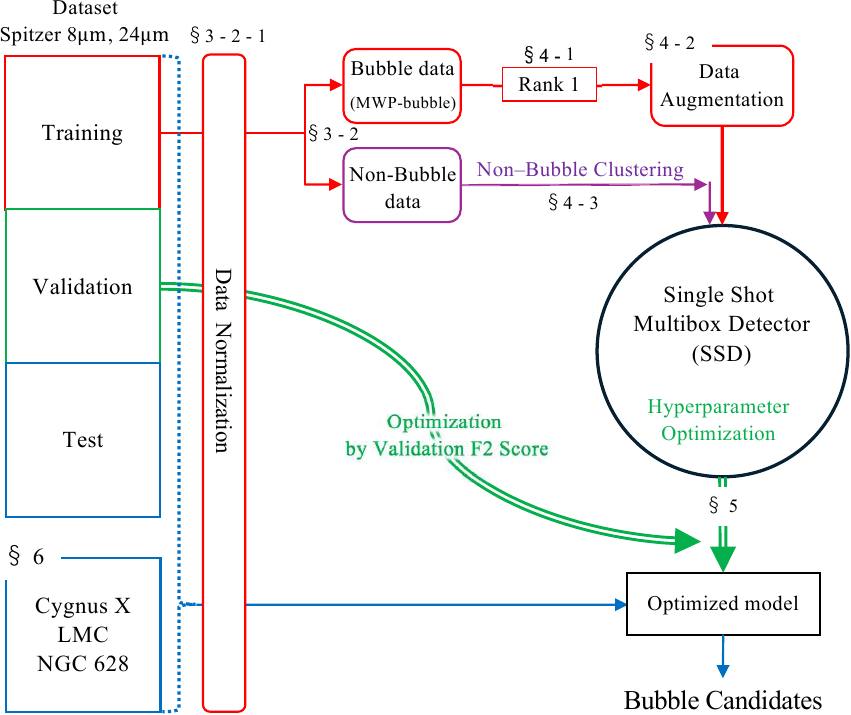}
    \end{tabular}
    \end{center}
    \caption
    {\label{fig:SSD_concept} 
    Processing flow of our deep learning model that can detect Spitzer bubbles. First, we divided the input data into training, validation, and test (left side, see Table~\ref{table:Training_Validation_Test}) to train the model and evaluate its performance. All data were normalized within the range of 0 to 1 as described in sub-subsection~\ref{sec:Processing_of_dataset}. Furthermore, training data are divided into two: Bubble and Non-Bubble data (subsection~\ref{sec:Dataset_used_for_training}). Second, to improve model performance (F2 score), Bubble data in training data were applied for bubble selection (subsection~\ref{sec:Selection_of_Spitzer_bubbles}) and data augmentation (subsection~\ref{sec:Effect_of_data_augmentation}), and hyperparameters of the model were Bayesian optimized. Finally, all data are input into the optimized model and the results output by the model are treated as bubble candidates (section~\ref{sec:Result_Discussion}). 
    {Alt text: A flow chart showing the processing flow of our deep learning model.}
    }
    \end{figure*}

Figure~\ref{fig:SSD_concept} shows the development procedure, with details of them described in section~\ref{sec:Details_of_data_optimization} and \ref{sec:Hyperparameter_optimization_our_model}.

\subsection{Comparison between SSD and the Ueda Model}
In this subsection, we show the advantages of SSD when compared to the Ueda Model in terms of inference time and the processing method of the results.

\subsubsection{Inference time}
Simple CNN-based classifiers such as the Ueda Model can only classify the presence or absence of objects in an image using confidence scores. Therefore, they determine the exact location of one Spitzer bubble in an image, by cropping a single image at various sizes and combining their inference results. Also, when there are multiple Spitzer bubbles in an image, they are processed in the same way. 

Similarly, since the original image is too large to be processed by SSD, our model also crops input images at various sizes for inference (see sub-subsection~\ref{sec:Creation_of_validation_data}). However, unlike classifiers, SSD in our model can output the location of objects in images and detect multiple objects simultaneously, thus it can use larger sizes for cropping. This capability allows our model to significantly reduce the number of cropped images needed to get the exact location and the inference time compared to the CNN model. For example, the inference time by GPU for a small region of 6 deg$^2$ is approximately 10 minutes for SSD with approximately 80,000 images to be inferred. On the other hand, the Ueda Model would use approximately 700 million images, assuming crop sizes of 23, 26, 28, ..., 2263, 2489, and 2738 pixels (calculated as $50 \times 1.1^x$, [$x$= -8$-$43]) and sliding-window strides of 1/10 of the crop sizes. As a result, it takes about 50 min for the inference by GPU. The inference time by GPU for SSD and CNN is not proportional to the number of cropped images because the Ueda Model has fewer CNN layers than SSD, resulting in a shorter inference time per image. In addition, including the time for processing image data, such as normalization and resizing, the total computation time for the Ueda Model exceeds several tens of hours. 

Based on this difference in inference time, SSD is more suitable than the Ueda Model for fast object detection.

\subsubsection{Post-processing after detection}
SSD is much simpler than the Ueda model in processing detection results. The Ueda Model calculates the position of Spitzer bubble using a probability cube created by connecting the probabilities of the cropped images at each crop size into a map and overlaying them. In this probability cube, a Spitzer bubble is determined if the number of connected voxels exceeding the probability threshold is greater than a specific value. However, such a complex process is time-consuming, and has several hyperparameters. In contrast, SSD can calculate the position of Spitzer bubbles without the complex analysis using probability cubes because SSD can output the location information. 

In addition, since SSD has a high detection accuracy \citep{liu2016ssd}, it can achieve the objective of faster speed without compromising accuracy, despite the different methods compared to the classifier. Thus, in this study, we used the SSD model.

\subsection{Single Shot multibox Detector (SSD)}
\label{sec:SSD}
    \begin{table}[t]
    \caption
    {\label{table:map_size_DBOX_type} 
    The resolutions of each source, the number of kinds of DBoxes, and the number of total DBoxes for each source. Source 1, 5, and 6 have four kinds of DBoxes and sources 2, 3 and 4 have six kinds of DBoxes, so source 1 has 5,776 ($38 \times 38 \times 4$) DBoxes and source 2 has 2,166 ($19 \times 19 \times 6$) DBoxes.}
    \begin{center}
    \renewcommand{\arraystretch}{1.3}
    \begin{tabularx}{\columnwidth}{@{\extracolsep{\fill}}ccrr}
    \toprule
             & \textbf{Resolution} & \textbf{Kinds} & \textbf{Total} \\ \midrule \midrule
    \textbf{source 1} & $38\times38$    & 4      & 5,776   \\
    \textbf{source 2} & $19\times19$    & 6      & 2,166   \\
    \textbf{source 3} & $10\times10$    & 6      & 600    \\
    \textbf{source 4} & $5\times5$      & 6      & 150    \\
    \textbf{source 5} & $3\times3$      & 4      & 36     \\
    \textbf{source 6} & $1\times1$      & 4      & 6      \\ \bottomrule   
    \end{tabularx}
    \end{center}
    \end{table}
SSD places 8,732 detection boxes of various sizes and shapes in the region of a single image (see the following paragraphs) and simultaneously performs classification and regression on all these boxes in a single inference. Then, the positions of the objects are determined from the offsets of those boxes obtained by the regression. In addition, SSD uses multiple feature maps with different resolutions for object detection, instead of a single feature map. By detecting smaller objects in feature maps of shallow layers and larger objects in feature maps of deeper layers, SSDs can detect objects of various sizes.

\begin{figure*}[htbp]
    \begin{center}
    \begin{tabular}{c} 
    \includegraphics[width=16cm]{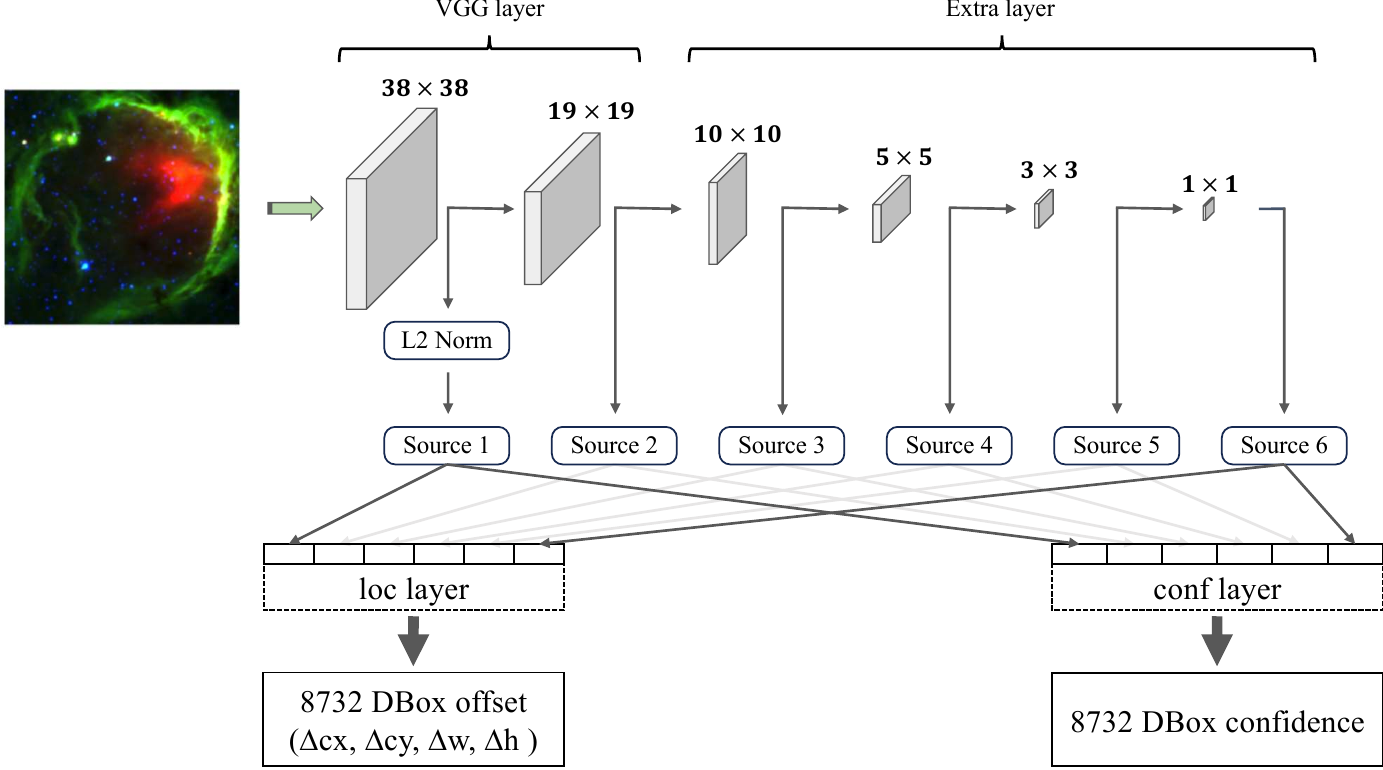}
    \end{tabular}
    \end{center}
    \caption
        { \label{fig:SSD_architecture} 
    The convolution flow and output of the SSD on which our model is based. The arrows indicate the sequence in which the data is convolved. Source 1--6 means feature maps generated by the convolution of the input image. Source 1 is the output after inputting the 38 $\times$ 38 $\times$ 512 feature map to L2 Norm, then all sources are input to the {\it loc} and {\it conf layers}. SSD can detect both large and small objects using the feature maps with 6 different resolutions. In this figure, only a single image is input, but it can be input in specified batch units. The input image is an example of Spitzer bubble with \SI{8}{\micro m} emission in green, and \SI{24}{\micro m} emission in red. For all subsequent images observed by the Spitzer Space Telescope, we will use the same scheme. See table~\ref{table:map_size_DBOX_type} for the size of each source and subsection~\ref{sec:SSD} for the DBox.
    {Alt text: A flow chart showing procedures for SSD to output 8,732 DBox offsets and confidence scores.}
    }
    \end{figure*}

SSD has four main components: {\it VGG layer}, {\it Extra layer}, {\it loc layer}, and {\it conf layer} (Figure~\ref{fig:SSD_architecture}).
The first {\it VGG layer} is based on VGG-16 (CNN with 16 layers) and the {\it loc} and {\it conf layer}s each consist of one convolution layer. These components work together as follows: The {\it VGG} and {\it Extra layer} are the feature extractors, producing 6 feature maps from the input image. The {\it loc layer} refines the positions and sizes of detected objects based on the feature map information. The {\it conf layer} assigns confidence scores to each detected object. Together, these layers help accurately localize and classify objects in the image.

The feature maps from source 1 to source 6 are extracted using the {\it VGG layer} and the {\it Extra layer}. Source 1 has features corresponding to small areas of an image, source 2 and source 3 have progressively larger areas in that order, and source 6 has features corresponding to large areas that represent the entire image. The resolutions of each source are shown in Table~\ref{table:map_size_DBOX_type}. Due to these four components and six feature maps, the SSD can accurately detect variously sized objects at high speed.
    
SSD estimates the exact position of objects using the relative offset values to the detection boxes. The fixed detection boxes are called Default-Boxes (DBoxes), and the boxes indicating the estimated position of the object with applied offsets are called Bounding-Boxes (BBoxes).
    
\begin{figure*}[htbp]
    \begin{center}
    \includegraphics[width=16cm]{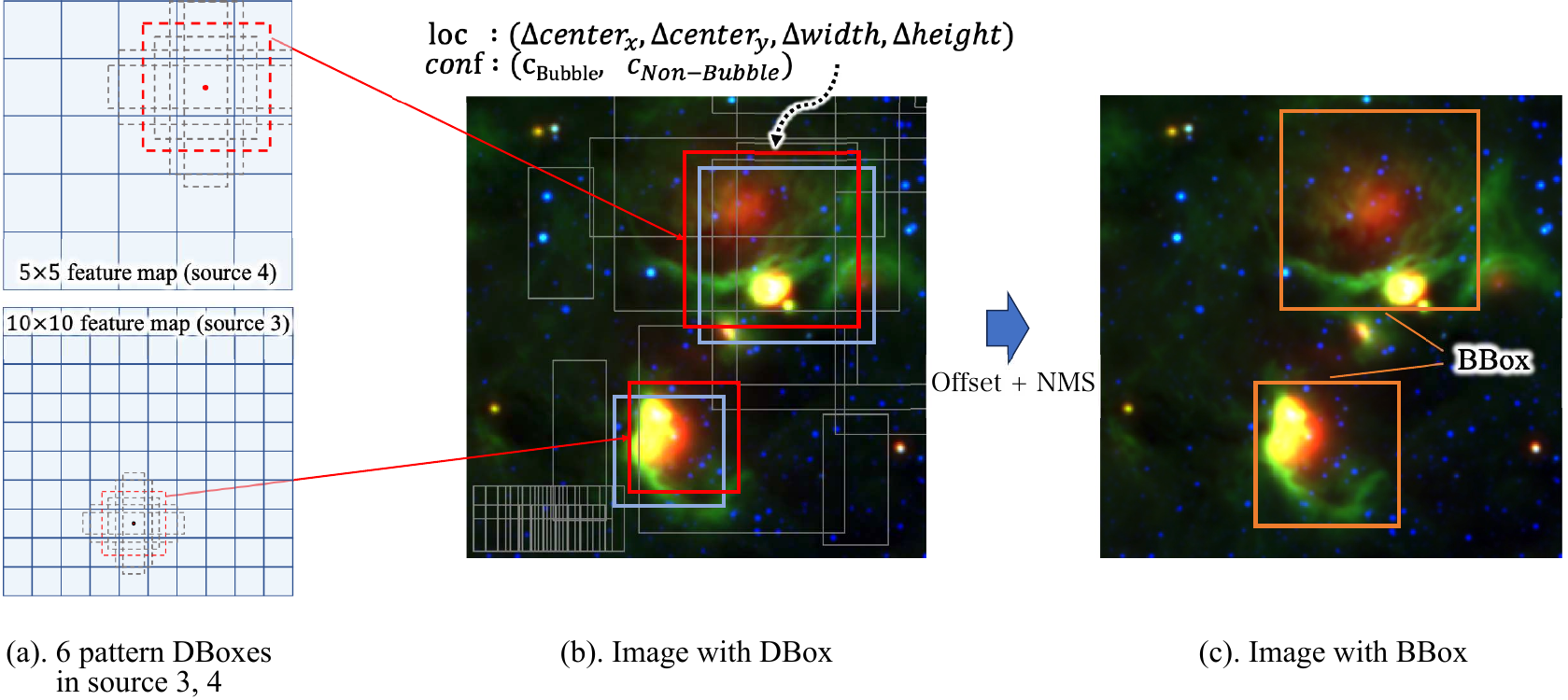}
    \end{center}
    \caption
    {\label{fig:DBox_BBox} 
     (a) Examples of feature maps and DBoxes for sources 3 and 4. The DBoxes exist in all cells, but we described only the DBoxes where objects are located in this figure. The DBox with the highest confidence is shown in red. In this case, only the DBoxes of sources 3 and 4 are taken as examples, but in reality, loc and conf are obtained for all DBoxes. See table 2 for the size of each source and the number of DBoxes. (b) The grey squares are dozens of DBoxes out of 8,732 DBoxes of different sizes and positions. The DBoxes in red are the DBoxes with the highest confidence of the object, and the blue color indicates the DBoxes with the next highest confidence. Although we only show several DBoxes for simplicity, there are more DBoxes around objects. (c) Because the objects and the DBoxes are slightly misaligned, SSD adjust them using the offset information and the equation defined by \citet{liu2016ssd}. Then, we use NMS to eliminate redundant BBoxes. The \SI{8}{\micro m} and \SI{24}{\micro m} emissions are shown in green and red, respectively.
     {Alt text: A diagram showing how BBoxes are made from the DBox of each source, output offsets, and confidence scores. The sequence from (a) to (b) to (c) illustrates the process from applying DBoxes to finalizing BBoxes.}
     }
    \end{figure*}

SSD detects objects using the six feature maps mentioned above and DBoxes corresponding to each feature map with different sizes and locations. The DBoxes are created by defining specific positions and sizes on the feature map grid (Figure~\ref{fig:DBox_BBox}a). 
There are a total of 8,732 DBoxes, consisting of 4 or 6 DBoxes with different aspect ratios for all 6 feature map cells (Table~\ref{table:map_size_DBOX_type}). SSD achieves superior execution speed compared to other methods by using multiple DBoxes for position estimation and class classification in a single inference.
    
Feature maps from source 1 to source 6 are input into the {\it loc} and {\it conf layers}, and convolution is performed once for each source. The {\it loc layer} outputs offset values ($\Delta$center$_{\text{x}}$, $\Delta$center$_{\text{y}}$, $\Delta$width, $\Delta$height) for each of 8,732 DBoxes. The {\it conf layer} outputs the confidence score for all object categories (c$_{\text{Bubble}}$, c$_{\text{Non-Bubble}}$) for each of 8,732 DBoxes. Then, the top 200 DBoxes with the highest c$_{\text{Bubble}}$ obtained from the {\it conf layer} are extracted. By substituting the coordinates of the DBoxes (center$_{\text{x}_\text{d}}$, center$_{\text{y}_\text{d}}$, width$_\text{d}$, height$_\text{d}$) and the offset values of the output into the equation defined by \citet{liu2016ssd} (Figure~\ref{fig:DBox_BBox}b), DBoxes are converted into the exact position coordinates of the objects (BBoxes). Furthermore, redundant BBoxes are eliminated by a method called Non-Maximum Suppression (NMS), leaving only one BBox per object (Figure~\ref{fig:DBox_BBox}c). In this study, the remaining BBoxes after NMS are treated as detected Spitzer bubbles.

Generally, SSD can detect multiple categories of objects (e.g., cars, people, and bicycles) at the same time; however, in this study, it is used to detect a single category, bubbles. In addition, We use two wavelength bands, \SI{8}{\micro m} and \SI{24}{\micro m}, for the detection of Spitzer bubbles. Therefore, the input was two-band data and the output was the background and Spitzer bubble classes.

\subsection{Dataset used for training}
\label{sec:Dataset_used_for_training}
    \begin{figure}[t]
    \begin{center}
    \begin{tabular}{c} 
    \includegraphics[width=8cm]{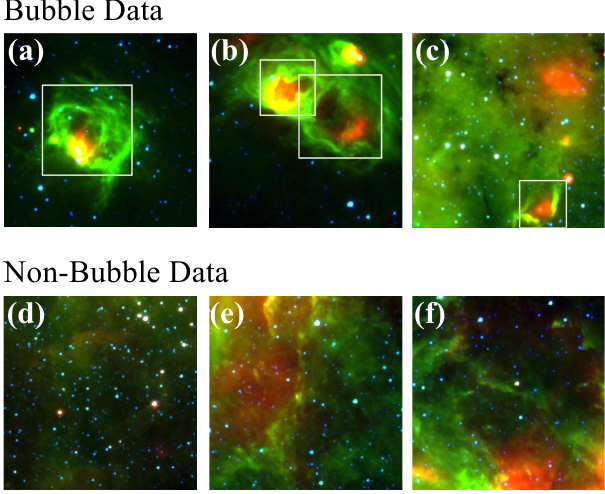}
    \end{tabular}
    \end{center}
    \caption
    {\label{fig:Bubble_NonBubble}
    Example of Bubble and Non-Bubble data; the white box in the Bubble data indicates the position of the MWP-Bubble. Both data have 300 $\times$ 300 pixels. The \SI{8}{\micro m} and \SI{24}{\micro m} emissions are shown in green and red, respectively.
    {Alt text: Three images each of Bubble data and Non-Bubble data.}
    }
    \end{figure}
The training data in this study are composed of two types: data with Spitzer bubbles (Bubble data) and data without Spitzer bubbles (Non-Bubble data) (Figure~\ref{fig:Bubble_NonBubble}). Bubble and Non-Bubble data consists of image data (data) and classes (labels) having the location information of the objects. The sequential procedure of updating the parameters of the SSD using all training data and evaluating the model with validation data is called epoch (see figure~6 in \cite{2022SPIE12189E..1QN}).

Bubble data (Figure~\ref{fig:Bubble_NonBubble}a, ~\ref{fig:Bubble_NonBubble}b, ~\ref{fig:Bubble_NonBubble}c) include Spitzer bubbles detected by the Milky Way Project (MWP-Bubble). The model developed in this study aims to detect the Spitzer bubble has a clear distribution with \SI{24}{\micro m} emission surrounded by \SI{8}{\micro m} emission.
Therefore, we used only the distinct MWP-Bubbles, corresponding to the bubbles categorized as Rank 1 (see subsection~\ref{sec:Selection_of_Spitzer_bubbles}), for the training and validation data. 
Additionally, the square area ratio of 175 bubbles shared by both MWP and CH06/07 shows that the area of MWP-Bubbles, calculated from the major axis (MajAxis), is approximately 1.7 times larger than the area of bubbles of CH06/07, calculated from the outer radius (R$_{\text{out}}$). In this study, to ensure the entire \SI{8}{\micro m} shell structure, we expanded the radii of MWP-Bubbles by a factor of 1.3, where 1.3 is the square root of 1.7. To ensure robust handling of partial bubble morphologies, we annotated bubbles on full-field Spitzer images before cropping. Partial bubbles were included in the training data only if more than 60\% of their structure was present within the cropped region. This approach minimizes false detections and enhances the model's performance.
The Bubble data used for training data are generated at each epoch, with a fixed random seed applied only at the beginning to ensure reproducibility. 

We also incorporated Non-Bubble data in training data to suppress false positives (backgrounds). Non-Bubble data were created by randomly cropping regions outside the areas of Rank 1 MWP-Bubbles, so Non-Bubble data is not expected to include the Spitzer bubble in the image (Figure~\ref{fig:Bubble_NonBubble}d, ~\ref{fig:Bubble_NonBubble}e, ~\ref{fig:Bubble_NonBubble}f). The role of Non-Bubble data is to make SSD learn the areas that are unrelated to the periphery of the Spitzer bubble as backgrounds. The Spitzer bubble exists only locally, and their area within the Milky Way is very small. If SSD is trained with only Bubble data, SSD can only learn the areas around the Spitzer bubble as backgrounds and cannot correctly learn the areas that are unrelated to the periphery of Spitzer bubbles as backgrounds. Therefore, including images with only backgrounds such as Non-Bubble data can avoid limiting the backgrounds to only the data surrounding the Spitzer bubble and suppress false positives. In this study, we applied Negative Sampling and Non-Bubble Clustering to the Non-Bubble data to train the model effectively (see subsection~\ref{sec:Non_Bubble_clustering}).

After cropping, both Bubble data and Non-Bubble data larger than 300 pixels are reduced to the resolution of 300 $\times$ 300 pixels and normalised using the method in subsection~{\ref{sec:Processing_of_dataset}}. Then, the data were resized to 300 $\times$ 300 pixels. Table~\ref{table:Training_Validation_Test} lists the regions used for the training, validation, and test. The training and validation data were randomly selected from FITS files excluding the test region.
    \begin{table}[t]
    \caption
    {\label{table:Training_Validation_Test}
    Areas of the training, validation, and test regions. After determining the test region, the training and validation regions were randomly determined. training, validation and test all with $|b| \leq 1^\circ$.}
    \begin{center}
    \renewcommand{\arraystretch}{1.3}
    \begin{tabularx}{\columnwidth}{@{\extracolsep{\fill}}lc}
    \toprule
    & \textbf{Galactic longitude range} \\ \midrule \midrule
    \textbf{Training}   & \begin{tabular}[c]{@{}c@{}}Area of $1^\circ \leq |l| \leq 65^\circ$ \\ excluding test and validation regions\end{tabular} \\
    \textbf{Validation} & \begin{tabular}[c]{@{}c@{}}$31.5^\circ \leq l \leq 34.5^\circ$, $37.5^\circ \leq l \leq 40.5^\circ$, \\ $46.5^\circ \leq l \leq 49.5^\circ$, $52.5^\circ \leq l \leq 55.5^\circ$\end{tabular} \\ 
    \textbf{Test}       & $10.5^\circ \leq l \leq 22.5^\circ$ \\ \bottomrule   
    \end{tabularx}
    \end{center}
    \end{table}

\subsubsection{Processing of dataset}
\label{sec:Processing_of_dataset}
    \begin{figure} [t]
    \begin{center}
    \begin{tabular}{c} 
    \includegraphics[width=8cm]{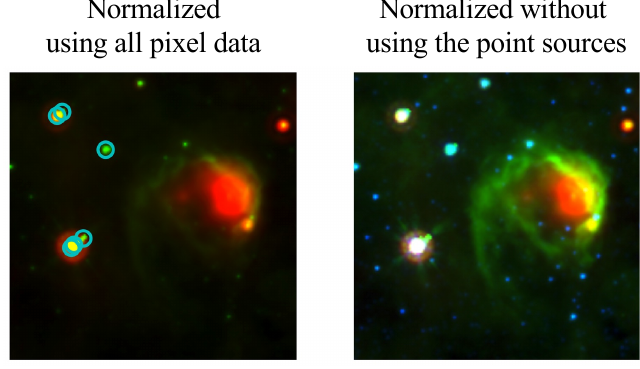}
    \end{tabular}
    \end{center}
    \caption
    { \label{fig:Normalization_processing} 
    Comparison of the same Spitzer bubble normalized using all pixel data (left), and normalized without using the point sources (right). The detected point sources are marked with cyan circles. The \SI{8}{\micro m} and \SI{24}{\micro m} emissions are shown in green and red, respectively.
    {Alt text: Figures showing the difference of brightness of \SI{8}{\micro m} emission. \SI{8}{\micro m} emission are visible in the right figure, but they have disappeared in the left figure.}
    }
    \end{figure}
In this study, to treat the values of \SI{8}{\micro m} and \SI{24}{\micro m} equally, the data were normalized for each channel to a range from 0 to 1 before being input into the model. The background level of the data obtained from the Spitzer Space Telescope was almost 0; however, for the JWST data, a specific value had to be subtracted to make the background level 0. The maximum value was set to three times the standard deviation above the mean intensity level of each target image. In this process, only for the \SI{8}{\micro m} data, the regions containing point sources are excluded, and the remaining data are normalized within the range of 0 to 1. The regions with point sources were assigned a value of 1 after normalization.

Figure~\ref{fig:Normalization_processing} compares the same Spitzer bubble, normalized using all pixels and normalized without using the point sources. As shown on the left side of Figure~\ref{fig:Normalization_processing}, when bright point sources are present, the \SI{8}{\micro m} distribution of the Spitzer bubble becomes close to 0 after normalization, making the bubble appear as if it is only a point source in the \SI{24}{\micro m} emission. To address this, point sources are identified and excluded using DAOStarFinder in photutils\footnote{$\langle$ \url{https://zenodo.org/records/12585239} $\rangle$}, based on their size and flux intensity. Specifically, point sources are defined as objects with a full width at half maximum (FWHM) smaller than or equal to 1.98 arcsec, matching the PSF of the \SI{8}{\micro m} observations. Additionally, to account for differences in flux intensity, we remove only point sources with intensities exceeding the mean plus three times the variance of the cropped data. This ensures that only the brightest point sources, which could significantly affect the normalization, are excluded while retaining small bubbles and other relevant structures for analysis.

Additionally, because all data are saved in the PNG format, they were converted to 256 gradations after the normalization. Then, when the data are input into the model, they are divided by 255 to fit within the range of 0 to 1.

\subsubsection{Creation of validation data}
\label{sec:Creation_of_validation_data}
\begin{figure} [t]
    \begin{center}
    \begin{tabular}{c} 
    \includegraphics[width=8cm]{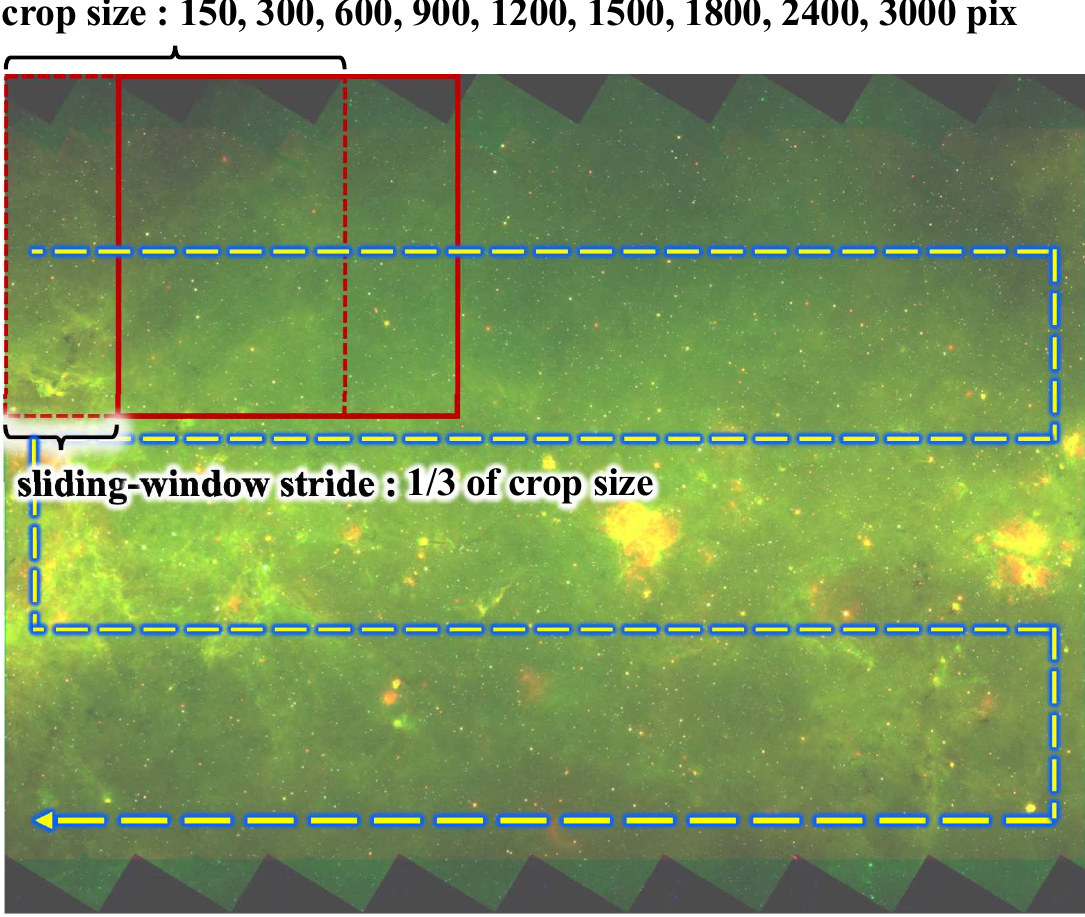}
    \end{tabular}
    \end{center}
        \caption
            { \label{fig:creation_of_ValidationData} 
        An example of how to crop data when creating validation data. The crop size is 150, 300, 600, 900, 1200, 1500, 1800, 2400, and 3000 pixels, and the data is cropped at sliding-window strides of 1/3 of the crop size. White dashed arrow indicates the direction in which the data is cropped. The \SI{8}{\micro m} and \SI{24}{\micro m} emissions are shown in green and red, respectively.
        {Alt text: Example of crop sizes and sliding-window stride on an image of the validation area.}
        }
    \end{figure}

Validation data and other data to be inferred are too large to be input directly into SSD, so these data are cropped into windows of various sizes and inferred.
We determined the crop sizes as half and multiples of 300 pixels, which is one edge of the SSD's input size. For example, crop sizes could be 150, 300, 600, or 900 pixels (Figure~\ref{fig:creation_of_ValidationData}). The sliding-window stride was set at 1/3 of the crop size to ensure finer image scanning. For instance, if the crop size is 300 pixels, the next cropping position would be at a stride of 100 pixels. This setup allows the SSD model to process the image more efficiently and accurately detect its features. The areas used for validation data are shown in Table~\ref{table:Training_Validation_Test}.
    \begin{figure} [t]
    \begin{center}
    \begin{tabular}{c} 
    \includegraphics[width=8cm]{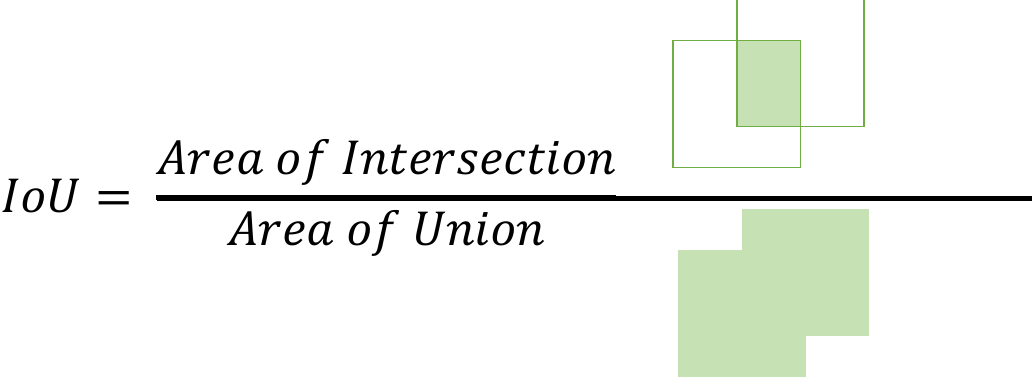}
    \end{tabular}
    \end{center}
        \caption
            { \label{fig:IoU} 
        IoU represents the percentage of overlapping Boxes. This figure is referenced from \citet{2022SPIE12189E..1QN}.
        {Alt text: IoU is calculated by dividing the area of intersection by the area of union.}
        }
    \end{figure}

\subsection{Loss}
In deep learning, loss refers to measuring how well or poorly a model's predictions match the true values. It's a numerical value that quantifies the error, with lower values indicating better performance. During training, models use backpropagation to adjust their parameters and minimize the loss by calculating gradients and updating weights.

The loss function for SSD is the sum of the confidence loss for class prediction and the location loss for bounding box regression. The location loss is calculated using the Smooth L1 Loss which is a loss function robust to outliers. The confidence loss is calculated using Cross-entropy Loss.

\begin{figure} [t]
    \begin{center}
    \begin{tabular}{c} 
    \includegraphics[width=8cm]{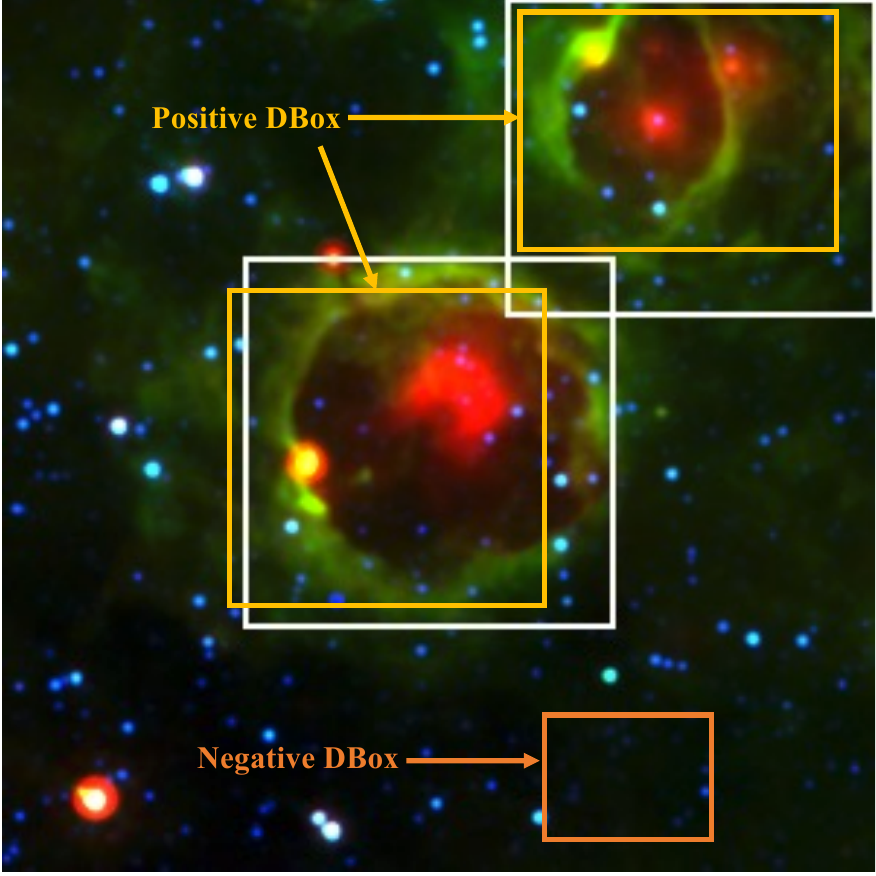}
    \end{tabular}
    \end{center}
    \caption
    { \label{fig:Positive_Negative_DBox} 
    Examples of Positive and Negative DBoxes. DBoxes with IoU$\geq$0.5 for the ground truth BBox (white boxes) are judged to be Positive DBoxes (yellow boxes). DBoxes with IoU$\leq$0.5 are judged to be Negative DBoxes (orange box).
    {Alt text: Example of negative and positive DBoxes on an image with two MWP-Bubbles.}
    }
    \end{figure}

SSD calculates the loss by dividing DBoxes into Positive DBox (DBox with Intersection over Union [IoU, Figure~\ref{fig:IoU}] $\geq$ 0.5 with the ground truth BBoxes [Spitzer bubble location information]) and Negative DBox (DBox without ground truth BBoxes with IoU $\geq$ 0.5) (Figure~\ref{fig:Positive_Negative_DBox}). For the Positive DBox, both location loss and confidence loss are calculated, while for the Negative DBox, only confidence loss is calculated. At this time, the number of Negative DBoxes becomes significantly larger than that of Positive DBoxes. Therefore, to avoid bias between Negative DBoxes and Positive DBoxes in training, the number of Negative DBoxes is limited to three times the number of Positive DBoxes. The Negative DBoxes with the highest confidence loss are selected. Additionally, because there are no Positive DBoxes in Non-Bubble Data (no ground truth BBoxes), the top 10 Negative DBoxes with the highest confidence loss were selected for training.

\subsection{Evaluation criteria}
In this study, we used precision, recall, and F2 score (a weighted harmonic mean of precision and recall, giving more weight to recall) as criteria to evaluate the maximum performance of the model. One of the purposes of this study is to find undetected Spitzer bubbles, and increasing precision may reduce the detection of new undetected Spitzer bubbles. Therefore, we used the F2 score, which emphasizes recall over precision, for performance evaluation. The equation (\ref{formula:Precision})--(\ref{formula:F2}) for calculating precision, recall, and the F2 score are as follows:
    \begin{eqnarray}
    \label{formula:Precision} 
precision = \frac{TP}{TP + FP}
    \end{eqnarray}
   \begin{eqnarray}
   \label{formula:Recall} 
recall = \frac{TP}{TP + FN}
   \end{eqnarray}
   \begin{eqnarray}
   \label{formula:F2} 
F2 \; score = \frac{5 \times precision \times recall}{4 \times precision + recall}
   \end{eqnarray}
TP (True Positive) represents the number of MWP-Bubbles correctly detected. FP (False Positive) indicates objects that were detected as Spitzer bubbles but are not MWP-Bubbles, while FN (False Negative) indicates objects that were detected as non-Spitzer bubbles but are cataloged as Spitzer bubbles by MWP.

Additionally, among the BBoxes that exceed the confidence threshold, those with IoU $\geq$ 0.5 are merged into the BBox with the highest confidence using Non-Maximum Suppression (NMS).

\section{Details of data optimization}
\label{sec:Details_of_data_optimization}
In this section, we introduce the effects of three optimizations to improve the performance of the model: 1) Selection of MWP-Bubbles, 2) Data Augmentation, and 3) Non-Bubble clustering. We illustrate the impact of each optimization by comparing the transitions in precision, recall, and F2 scores. In subsection~\ref{sec:Selection_of_Spitzer_bubbles}, \ref{sec:Effect_of_data_augmentation}, and~\ref{sec:Non_Bubble_clustering}, we conducted experiments with default values of mini-batch size = 8, learning rate = 1$\times10^{-4}$ and weight decay = 1$\times10^{-4}$ (see section~\ref{sec:Hyperparameter_optimization_our_model}).
In addition, SSD utilizes Negative Sampling, a form of random sampling, for Non-Bubble data to facilitate effective learning.
    \begin{figure*} [t]
    \begin{center}
    \begin{tabular}{c} 
    \includegraphics[width=16cm]{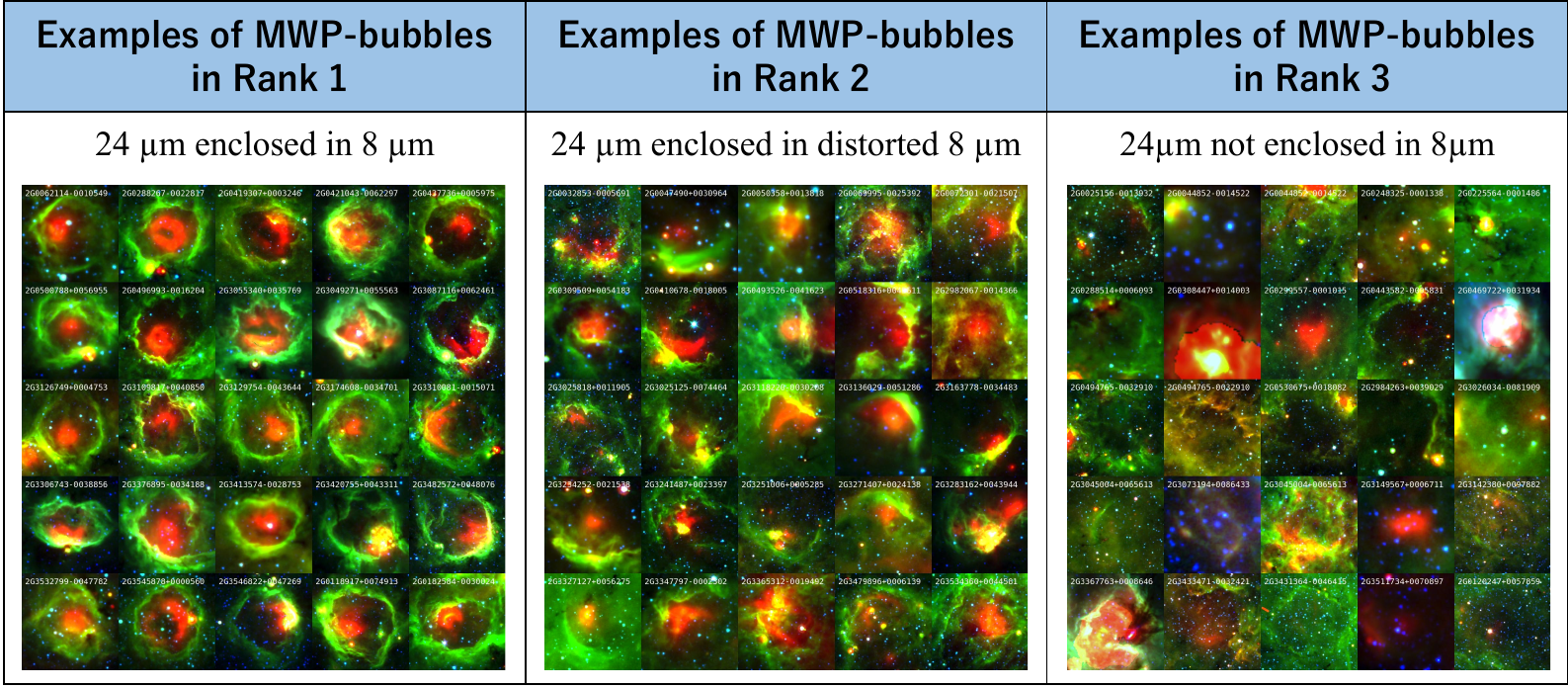}
    \end{tabular}
    \end{center}
    \caption
    { \label{fig:MWP_Rank} 
    A part of three ranks of MWP-Bubbles, Rank 1 (\SI{8}{\micro m} encompassing \SI{24}{\micro m}, 634 bubbles), Rank2 (distorted \SI{8}{\micro m} encompassing \SI{24}{\micro m}, 952 bubbles), and Rank3 (no encompassing, 815 bubbles). The name of the used MWP-Bubble is noted in the top left-hand corner of each image. The \SI{8}{\micro m} and \SI{24}{\micro m} emissions are shown in green and red, respectively.
    {Alt text: MWP-Bubble Examples by each rank.}
    }
    \end{figure*}
    \begin{figure*} [t]
    \begin{center}
    \begin{tabular}{c} 
    \includegraphics[width=16cm]{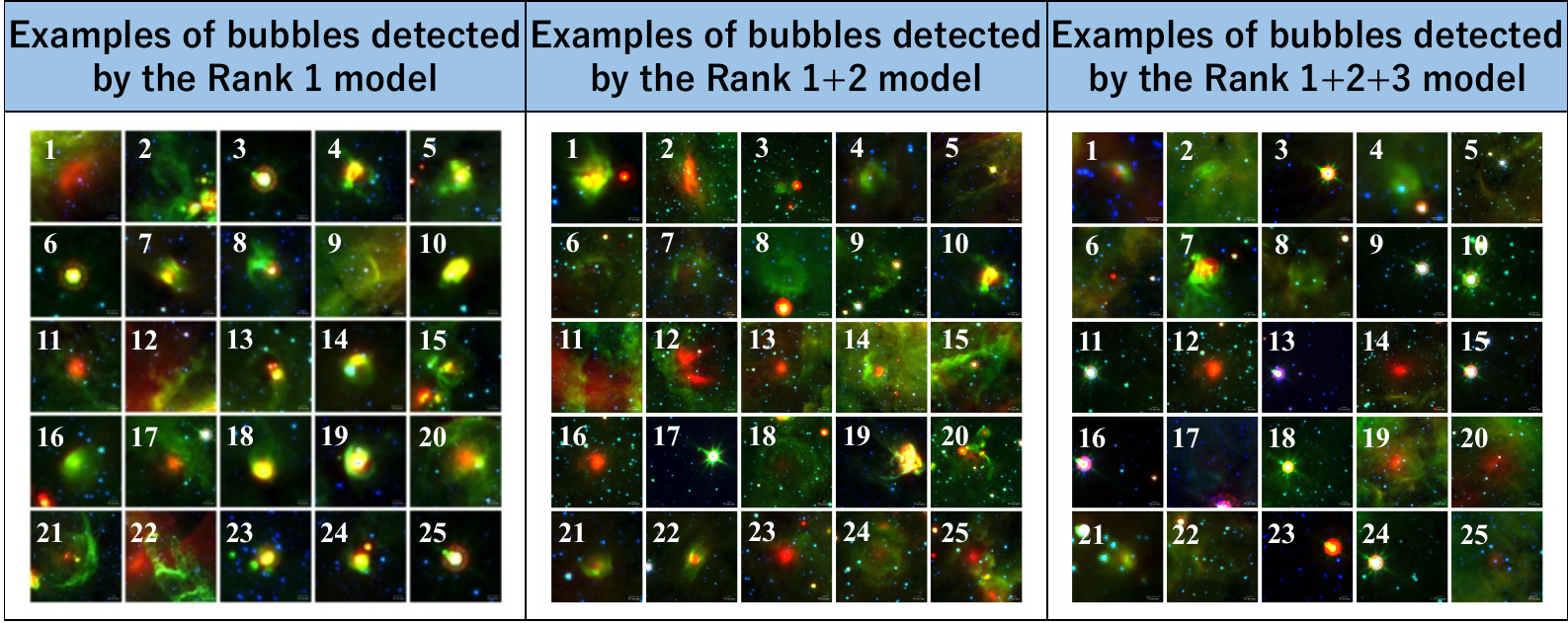}
    \end{tabular}
    \end{center}
    \caption
    {\label{fig:Result_Selection_of_Spitzer_bubbles} 
     Comparison of newly detected objects as Spitzer bubble in the test region when changing the ranked MWP-Bubble used in training data. Id = 4, 5, 7, 8, 10, 13, 14, 15, 16, 17, 18, 19, 20, 21, 23, and 24 (16/25) for Rank 1. Id = 1, 4, 7, 10, 12, 14, 15, 19, 20, 21, 22, 24, and 25 (13/25) for Rank 1+2. Id = 1, 7, and 19 (3/25) for Rank 1+2+3 captures features of objects formed by the radiation of high-mass stars. The \SI{8}{\micro m} and \SI{24}{\micro m} emissions are shown in green and red, respectively.
     {Alt text: Examples of detection by each model.}
     }
    \end{figure*}

\subsection{Selection of Spitzer bubbles}
\label{sec:Selection_of_Spitzer_bubbles}
The MWP-Bubbles include many objects with unclear \SI{8}{\micro m} shell structures and \SI{24}{\micro m} distributions. These objects can obscure the criteria by which the model detects Spitzer bubbles and negatively impact the detection of clear Spitzer bubbles. To understand how data ambiguity affects model accuracy, we ranked MWP-Bubbles into three categories and trained the model using training and validation data with different ranks (Rank 1, Rank 1+2, Rank 1+2+3). We applied these three models to the test region and determined the optimal rank for training and validation data based on the accuracy of newly detected bubbles. 

In this study, we classified the MWP-Bubbles used into three patterns: Rank 1, Rank 2, and Rank 3, as shown in Figure~\ref{fig:MWP_Rank}. Rank 1 includes bubbles where \SI{8}{\micro m} encloses \SI{24}{\micro m}. Rank 2 includes bubbles where a distorted \SI{8}{\micro m} encloses \SI{24}{\micro m}. Rank 3 includes bubbles where \SI{8}{\micro m} does not enclose \SI{24}{\micro m}. The numbers of bubbles identified as Rank 1, Rank 2, and Rank 3 were 634, 952, and 815, respectively. The total number of bubbles classified as Rank 1, Rank 2, and Rank 3 is 2,401, excluding MWP-Bubbles that span multiple FITS files. We evaluated the performance of the model by changing the MWP-Bubbles used in training and validation data to Rank 1, Rank 1+2, and Rank 1+2+3, using the results of the test region. Because the validation data also uses ranked MWP-Bubbles, each model has different validation data. Therefore, the comparisons in terms of precision, recall, and the F2 score are only for reference. The performance of each model is compared using inference results in the test region. In the test region, there are 84 Rank 1 MWP-Bubbles, 111 Rank 2 MWP-Bubbles and 91 Rank 3 MWP-Bubbles.

We created training and validation data in three patterns: using only Rank 1, Rank 1+2, and Rank 1+2+3 MWP-Bubbles. We trained three models with these three patterns of data and applied the three trained models to the test region. As a result, the number of objects newly detected as bubbles in the test region was 193 for the model trained with Rank 1, 625 for the model trained with Rank 1+2, and 875 for the model trained with Rank 1+2+3. 
\setlength{\parskip}{0.7mm}

Figure~\ref{fig:Result_Selection_of_Spitzer_bubbles} displays 25 randomly selected objects detected by each model. It is clear that the models trained with Rank 1+2 and Rank 1+2+3 MWP-Bubbles as training data contain a significant number of false positives, i.e., structures that are evidently not associated with massive stars.
Therefore, we conclude that the inclusion of MWP-Bubbles with unclear \SI{8}{\micro m} shell structures and \SI{24}{\micro m} distributions in the training data increases false positives. In subsequent experiments, we used only Rank 1 MWP-Bubbles for training and validation data.

\subsection{Effect of data augmentation}
\label{sec:Effect_of_data_augmentation}
\begin{figure} [t]
    \begin{center}
    \begin{tabular}{c} 
    \includegraphics[width=8cm]{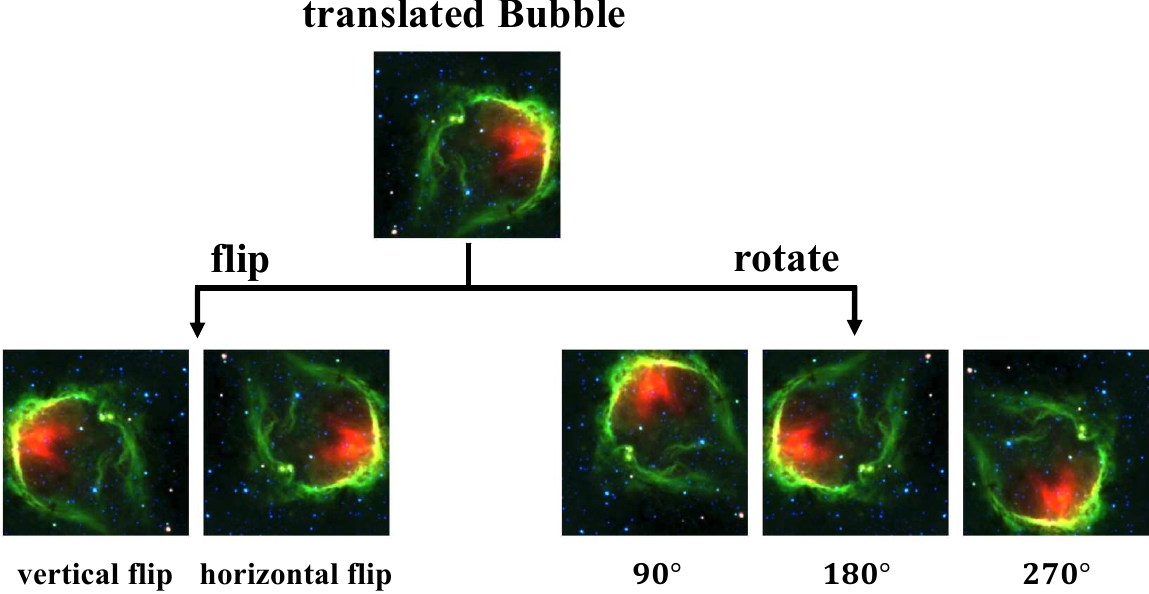}
    \end{tabular}
    \end{center}
    \caption
        { \label{fig:example_DA} 
     Example of data augmentation. Rank 1 MWP-Bubbles were translated and augmented using five patterns of flipping and rotation. The augmented data included upside-down and left-right flipped images, as well as images rotated by \ang{90}, \ang{180}, and \ang{270}. An additional experiment with further rotation angles (\ang{45}, \ang{135}, \ang{225}, and \ang{315}) resulted in an F2 score of 0.619, which is comparable to the score without these additional angles (F2 score = 0.666). Based on these findings, we limited rotations to \ang{90}, \ang{180}, and \ang{270}. {Alt text: Upside-down and left-right flipped images of the MWP-Bubble after it has been moved, and images rotated by \ang{90}, \ang{180}, and \ang{270}.}
    }
    \end{figure}

Data augmentation (DA) is expected to improve model performance by expanding the amount of data through image processing including rotation and flipping, thereby preventing overfitting. We attempted to improve performance with DA, because there are only 647 Rank 1 MWP-Bubbles. In this study, we used translation, rotation, and flipping as data augmentation (DA) techniques. Rotation and flipping were applied to Spitzer bubbles that had been translated beforehand, as shown in Figure~\ref{fig:example_DA}. While numerous other augmentation methods, such as cut-out \citep{DBLP:journals/corr/abs-1708-04552}, mix-up \citep{zhang2018mixup}, and GAN-based image generation \citep{NIPS2014_5ca3e9b1}, are available, we focused on these conventional techniques to ensure a balanced and effective training process. Future work could explore the potential benefits of incorporating more advanced augmentation strategies.
In the translation process, a bubble is cropped from the data and randomly positioned within the image, ensuring it remains fully contained within the image boundaries. For each epoch, the cropping size of the bubble is randomly chosen, ranging from 1.3 to 6 times the actual size of the bubble. The upper limit of 6 times the size ensures that the bubble remains recognizable within a 300 $\times$ 300 image. 
We updated the Bubble data created in this way every epoch with a fixed random seed to improve the generalization performance of the model (as explained in subsection~\ref{sec:Dataset_used_for_training}).

    \begin{figure} [t]
    \begin{center}
    \begin{tabular}{c} 
    \includegraphics[width=8cm]{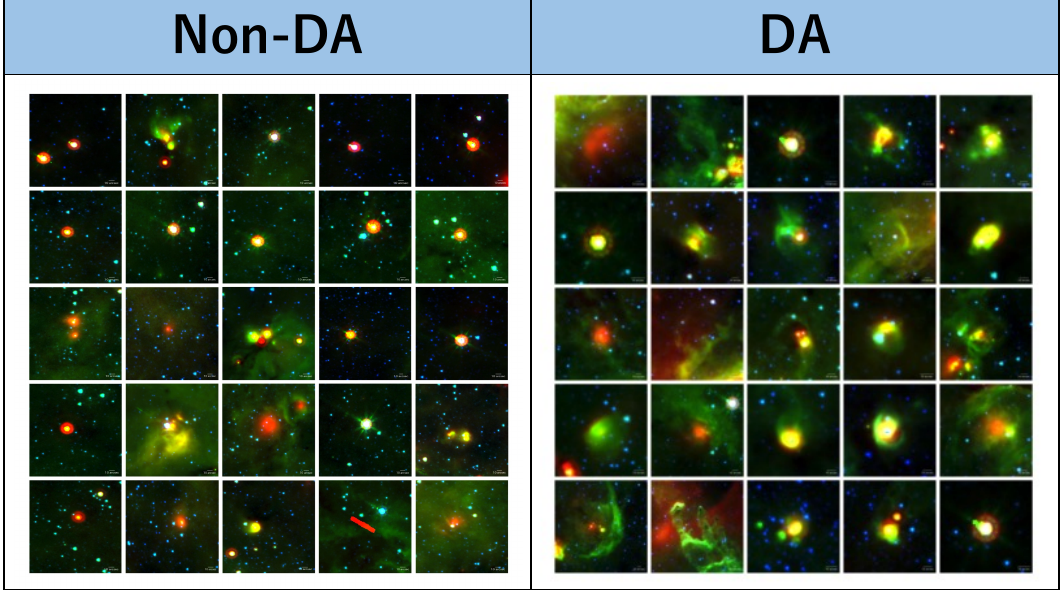}
    \end{tabular}
    \end{center}
    \caption
    { \label{fig:Result_DA} 
    Comparison of the 25 randomly selected objects that were newly detected as Spitzer bubbles by the two models trained on training data with data augmentation (DA) and training data without DA (Non-DA). The \SI{8}{\micro m} and \SI{24}{\micro m} emissions are shown in green and red, respectively.
    {Alt text: The detection results by each model.}
    }
    \end{figure}

To measure the performance improvement with DA, we compared the precision, recall, and F2 score using the test results of the model trained with and without DA. Figure~\ref{fig:Result_DA} shows a comparison of 25 randomly selected newly detected objects classified as Spitzer bubbles when evaluating the test region. The model trained without DA detected many point sources, whereas the model trained with DA detected many objects that could be considered Rank 1 or 2 MWP-Bubbles. In this region, the precision, recall, and F2 score with DA were 0.293, 0.976, and 0.666, respectively, while the precision, recall, and F2 score without DA were 0.0596, 0.833, and 0.232, respectively. Based on these results, we conclude that training with DA increased the recall of Rank 1 MWP-Bubbles and effectively suppressed false positives due to the enhanced background patterns in the Bubble data. Therefore, in subsequent experiments, we applied DA to the Bubble data in addition to selecting Spitzer bubbles for the training data.

\subsection{Negative Sampling and Non-Bubble clustering}
\label{sec:Non_Bubble_clustering}
    \begin{figure*} [t]
    \begin{center}
    \begin{tabular}{c} 
    \includegraphics[width=15cm]{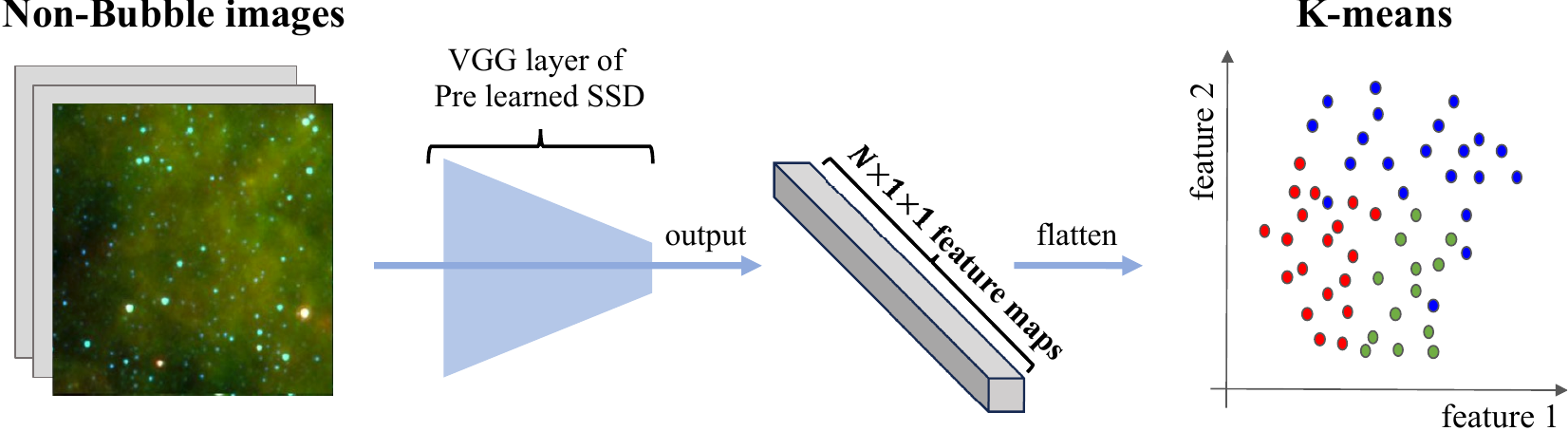}
    \end{tabular}
    \end{center}
    \caption
        { \label{fig:Non_Bubble_clustering} 
    The method of clustering Non-Bubble data (300 $\times$ 300 pixel) using feature maps ($N$ $\times$ 1 $\times$ 1). First, we created Non-Bubble data. Second, Non-Bubble data was compressed down to 1$\times$1 using the {\it VGG layer} of pre-learned SSD. Finally, they were clustered by $k$-means after being flattened. The $k$-means is a non-hierarchical clustering method that divides data into $k$ clusters (\cite{macqueen1967some}). The figure shows an example when classified into three classes.
    {Alt text: A diagram illustrating the method of clustering Non-Bubble data.}
    }
    \vspace{-13pt}
    \end{figure*}

    \begin{figure*} [t]
    \begin{center}
    \begin{tabular}{c} 
    \includegraphics[width=15cm]{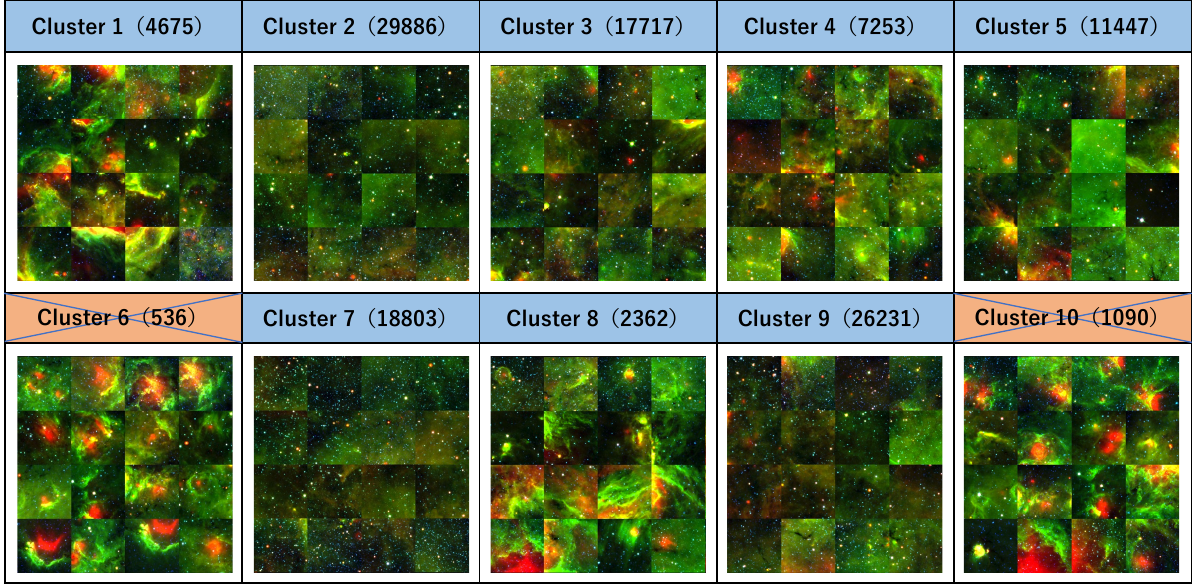}
    \end{tabular}
    \end{center}
    \caption
    {\label{fig:Result_NonRing_CL} 
    The results of clustering the Non-Bubble data by the method shown in Figure~\ref{fig:Non_Bubble_clustering}. The number of images for cluster 1--10 was 4,675, 29,886, 17,717, 7,253, 11,447, 536, 18,803, 2,362, 26,231, and 1,090 respectively.  Due to the large number of images, only 16 of each class were randomly excerpted. This result shows many images with no structures, such as clusters 2, 7, and 9. Because the number of images in each cluster differs greatly, it is clear that the SSD cannot uniformly learn the characteristics of each cluster using simple random sampling from all Non-Bubble data. Clusters 6 and 10 with Spitzer bubble-like structure were excluded from Non-Bubble data (marked with a cross). By learning for each cluster, SSD can efficiently learn images of clusters 1 and 8, which have structures that are prone to false positives.
    {Alt text: Figures showing the number and image samples of each cluster as a result of clustering the Non-Bubble data into 10 clusters.}
    }
    \vspace{-17pt}
    \end{figure*}

Here, we select the same number of Non-Bubble objects as Data Augmented Bubbles for training. Randomly selected Non-Bubble data may result in the inclusion of Spitzer bubbles that have not been previously detected, which could degrade the training performance. Additionally, most Non-Bubble images contain only stars statistically, as shown in Figure~\ref{fig:Bubble_NonBubble}d. Therefore, it is important to include infrared structures that are not Spitzer bubbles, such as infrared ridges or cores, as Non-Bubble objects in the negative data, in order to prevent these structures from being misidentified as Spitzer bubbles.


To address this issue, we first randomly selected 120,000 Non-Bubble data samples from all regions and applied a clustering-based Negative Sampling method, as illustrated in Figure~\ref{fig:Non_Bubble_clustering}. Using $k$-means clustering, we divided the Non-Bubble data into 10 clusters. This method proved to be the most effective for clustering Non-Bubble data, including undetected Spitzer bubbles. The breakdown of these 10 clusters is shown in Figure~\ref{fig:Result_NonRing_CL}.
Among these clusters, clusters 6 and 10 were excluded because they were found to likely contain Spitzer bubbles. For the remaining clusters, the number of images per cluster was balanced through Negative Sampling, ensuring that each cluster contributed an equal number of samples. This balancing approach helped optimize the training process for Non-Bubble data.
Of particular importance is that clusters 1, 4, 5, and 8 include complex infrared structures, such as filaments and ridges, which are not related to Spitzer bubbles. By including these structures in the Negative Sampling process, we ensured that the training dataset accounted for diverse non-bubble features, thereby enhancing the model's robustness.

To measure the performance, we compared precision, recall, and F2 score with the test results of the model trained with and without Non-Bubble clustering. In the test region, precision, recall, and F2 score with Non-Bubble clustering were 0.419, 0.893, and 0.728, respectively, and precision, recall, and F2 score without Non-Bubble clustering were 0.293, 0.976, and 0.666, respectively.
    
For the purpose of creating pure Non-Bubble data, it is also possible to apply a pre-trained model to all regions and recreate Non-Bubble data from regions outside MWP-Bubbles and newly detected bubbles, in addition to the clustering method. However, this method is more time consuming and can not efficiently learn the Non-Bubble data with various features; therefore, we adopted the clustering method.

From these results, we conclude that the performance of the model improves by selecting Spitzer bubbles, performing Data Augmentation, adding infrared structures not related to the bubbles, and removing undetected bubbles through clustering Non-Bubble data in the training data.
    \begin{figure} [htbp]
    \begin{center}
    \begin{tabular}{c} 
    \includegraphics[width=8cm]{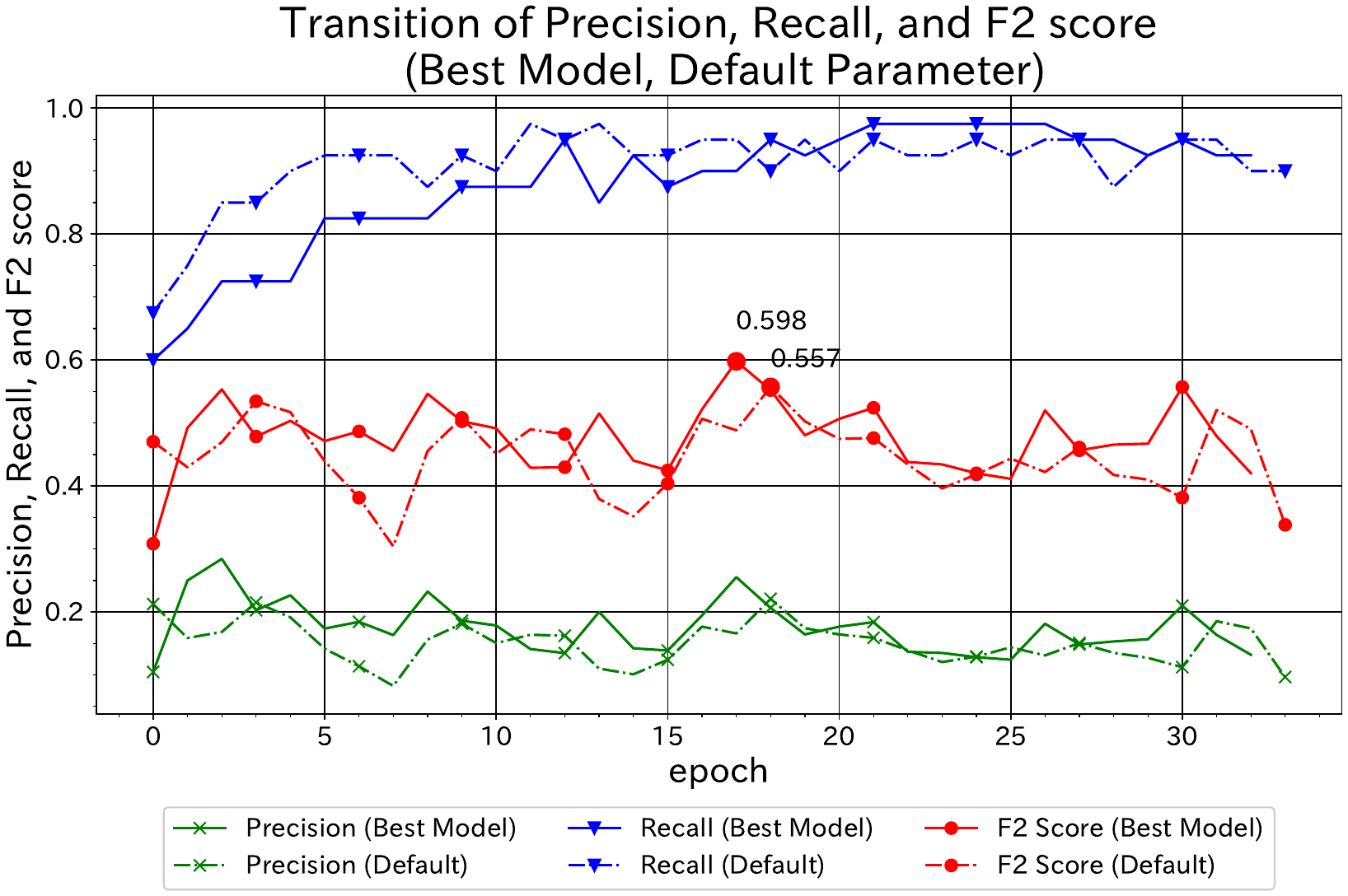}
    \end{tabular}
    \end{center}
    \caption
    { \label{fig:Comparison_F2_HyperResearch} 
    Comparison of precision, recall, and F2 scores of the best and default models. The x-axis represents the number of epochs, while the y-axis represents the percentage of precision, recall, and F2 score ranging from 0 to 1.
    {Alt text: Line graph showing the transition of precision, recall, and F2 score for each epoch of the best and default model.}
    }
    \end{figure}

\section{Hyperparameter optimization}
\label{sec:Hyperparameter_optimization_our_model}
    \begin{table}[t]
    \caption
    {\label{table:Hyperparameter} 
    Loss and mini-batch hyperparameters to be explored. The learning rate and weight decay were explored in the range of 0.00001 to 0.1, and the mini-batch was explored in the range of 2, 4, 8, 16, and 32.
    }
    \begin{center}
    \renewcommand{\arraystretch}{1.3}
    \begin{tabularx}{\columnwidth}{@{\extracolsep{\fill}}lcc}
    \toprule
    \multicolumn{3}{c}{\textbf{Loss Hyperparameter}}       \\ \midrule \midrule
    Type               & min              & max   \\
    Learning rate      & 0.000001         & 0.001 \\
    Weight decay       & 0.000001         & 0.001 \\ \toprule
    \multicolumn{3}{c}{\textbf{Mini-Batch} Hyperparameter} \\ \midrule \midrule
    Type               & Batch size       &       \\
    Bubble mini-batch  & 2, 4, 8, 16, 32  &       \\   \bottomrule
    \end{tabularx}
    \end{center}
    \vspace{-8pt}
    \end{table}
    
    \begin{figure*} [t]
    \begin{center}
    \begin{tabular}{c} 
    \includegraphics[width=17.3cm]{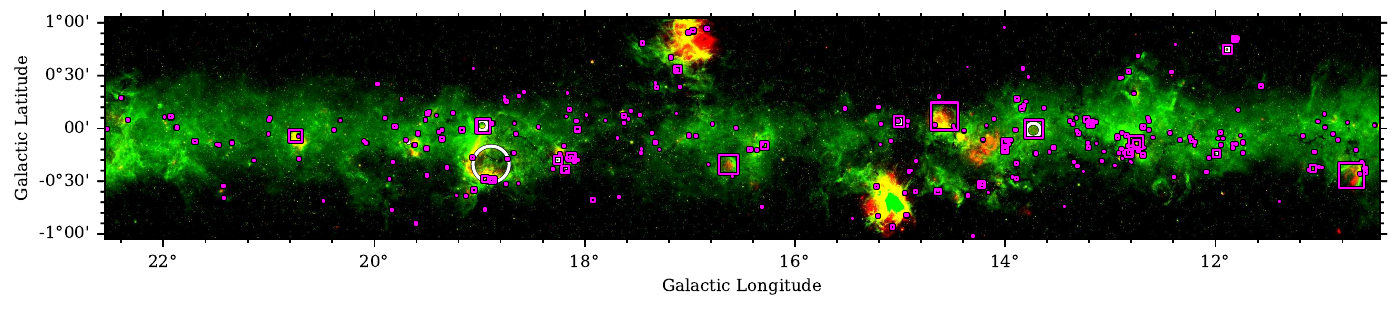}
    \end{tabular}
    \end{center}
    \caption
    {\label{fig:Test_detection_map} 
    The test region ($10.5^{\circ} \leq l \leq 22.5^{\circ}$, $-1^{\circ} \leq b \leq 1^{\circ}$) with 289 objects detected by our model as Spitzer bubbles (magenta squares) and Rank 1 MWP-Bubble (white circles). The \SI{8}{\micro m} and \SI{24}{\micro m} emissions are shown in green and red, respectively.
    {Alt text: A map of the test region showing 289 objects detected by our model as Spitzer bubbles, with Rank 1 MWP-Bubbles marked.}
    }
    \vspace{-12pt}
    \end{figure*}
    
In SSD, some hyperparameters, such as Bubble mini-batch size, learning rate, and weight decay, need to be predetermined before training. We attempted to improve the performance of the model by optimizing the hyperparameters using Bayesian optimization which is one of the hyperparameter optimization methods. In this study, we searched hyperparameters within the range shown in Table~\ref{table:Hyperparameter} using the Weights \& Biases Sweeps. We searched them 40 times and found that the F2 score tended to be higher when the learning rate and weight decay were below 0.0001 and the Bubble mini-batch size was 32. The values of each parameter when the F2 score to validation data was highest at 0.598 were as follows: learning rate is 7.7461$\times10^{-5}$, weight decay is 8.3171$\times10^{-5}$, and Bubble mini-batch is 32. Figure~\ref{fig:Comparison_F2_HyperResearch} compares the changes in F2 score, recall, and precision between the default hyperparameters and the best parameters. The precision increased overall, and although the increase in recall was smaller than the default parameters, it eventually reached the same level. In section~\ref{sec:Result_Discussion}, we use the model obtained with the best parameters.\footnote{The python codes of our model are available at  $\langle$\url{https://github.com/ninpei7114/galactic_bubble}$\rangle$.}

\section{Result \& Discussion}
\label{sec:Result_Discussion}
    
\subsection{Detection of Spitzer Bubble}
In this section, we show the performance and validity of our model. First, we apply the model to the test region and compare the result with the  MWP-Bubbles. Then, we extend the application to the training and validation regions. Last, we examine the validity of our model for detecting Spitzer bubbles in other data in three regions: Cygnus X, the LMC, and NGC 628. We show here the characteristics of the detected bubbles. For the inferences, we used crop sizes of 100, 150, 300, 600, 900, 1200, 1500, 1800, 2400, and 3000 pixels, and the sliding-window stride was set at 1/3 of the cropping size.

\subsubsection{Application to test region}
\label{sec:Application_to_test_region}
    \begin{table}[t]
    \caption
    {\label{table:confusion_matrix_test} 
    Confusion matrix for the test region. The Rank 1 MWP-Bubbles in the test region is 84, while our model detected 289 objects.
    }
    \begin{center}
    \renewcommand{\arraystretch}{1.3}
    \begin{tabularx}{\columnwidth}{@{\extracolsep{\fill}}|c|c|c|} \hline
    \diagbox{MWP}{Predicted} & Bubble & Non-Bubble \\ \hline
    Bubble & 82 & 2 \\ \hline
    Non-Bubble & 207 & - \\ \hline
    \end{tabularx}
    \end{center}
    \end{table}
    \begin{figure} [t]
    \begin{center}
    \begin{tabular}{c} 
    \includegraphics[width=8cm]{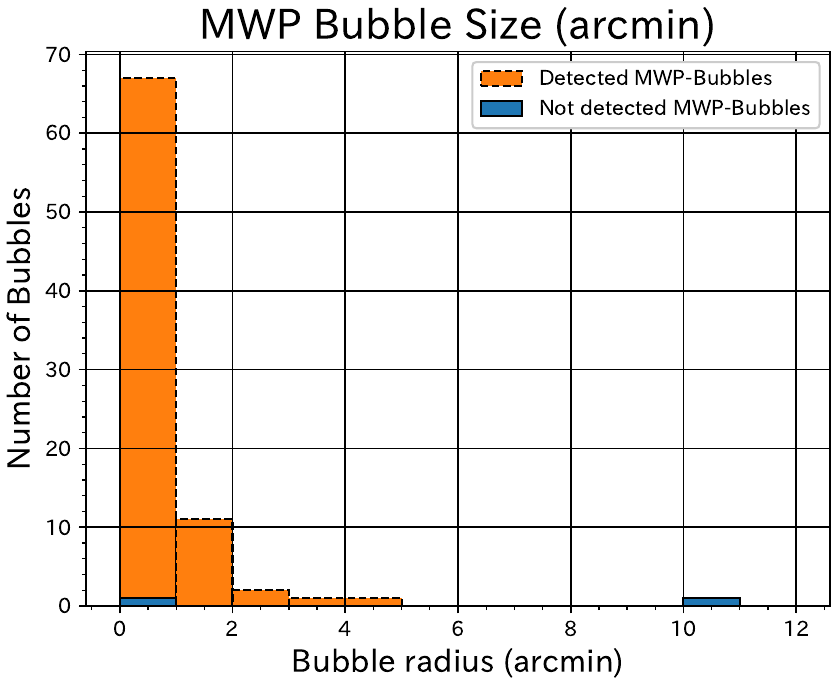}
    \end{tabular}
    \end{center}
    \caption
        { \label{fig:Test_Size_comparison_of_MWP_Bubble} 
    Size comparison of Rank 1 MWP-Bubble that could and could not be detected by our model.
    {Alt text: A histogram showing the size of Rank 1 MWP-Bubble that could and could not be detected by our model.}
    }
    \end{figure}
    \begin{figure*} [t]
    \begin{center}
    \begin{tabular}{c} 
    \includegraphics[width=16cm]{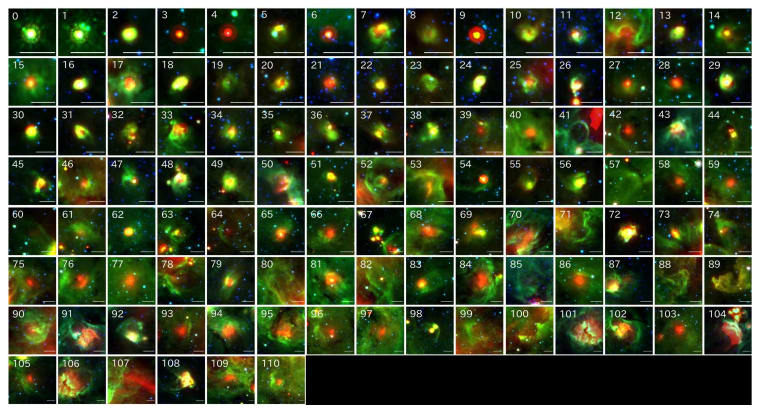}
    \end{tabular}
    \end{center}
    \caption
        { \label{fig:Test_each_image} 
    The 111 newly detected objects as Spitzer bubbles by our model in the test region, sorted by size. The scale bar is 40$''$. The \SI{8}{\micro m} and \SI{24}{\micro m} emissions are shown in green and red, respectively.
    {Alt text: Summary of images showing 111 newly detected objects in the test region.}
    }
    \vspace{-3pt}
    \end{figure*} 
We applied the model to the test region, which was not used for training and validation. The test region is set to $10.5^{\circ} \leq l \leq 22.5^{\circ}$, $-1^{\circ} \leq b \leq 1^{\circ}$ (Table~\ref{table:Training_Validation_Test}), and contains 286 MWP-Bubbles (Rank 1: 84, Rank 2: 111, Rank 3: 91). The model evaluated the test region and detected 289 objects (inference time, 23 min). Figure~\ref{fig:Test_detection_map} shows the detected objects in magenta and Rank 1 MWP-Bubbles in white. About 80\% of the detected objects are located within galactic latitudes $-0.5^{\circ} \leq b \leq 0.5^{\circ}$, and various sizes of objects can be seen. Table~\ref{table:confusion_matrix_test} compares the detected objects with Rank 1 MWP-Bubbles. Out of the 84 Rank 1 MWP-Bubbles, 82 were detected, resulting in a very high detection rate of 98\% for Rank 1 MWP-Bubbles.

Figure~\ref{fig:Test_Size_comparison_of_MWP_Bubble} shows a histogram of the sizes of bubbles that could and could not be detected among the Rank 1 MWP-Bubbles. The two undetected bubbles were \timeform{0'.34} (=17 pixels) and \timeform{10'.34} (=517 pixels) in size, indicating that both the smallest and largest bubbles were missed. The detection rate by each MWP-Bubble rank is 98\% for Rank 1, 63\% for Rank 2, and 22\% for Rank 3. Compared to all MWP-Bubbles, the detection rate was 60\%. The 40 \% of MWP-Bubbles not detected by our model had distorted shapes, and most of them may be objects that have had some time since bubble formation, have complex environments, or are not associated with high-mass star formation.

Figure~\ref{fig:Test_each_image} shows 111 objects that were newly detected as Spitzer bubbles by the model. At least 50\% of the newly detected objects can be regarded as Rank 1 and 2 with \SI{8}{\micro m} encompassing \SI{24}{\micro m}, such as id = 7, 10, 20, 25, 48 and 90, indicating the effectiveness of the optimization performed in sections~\ref{sec:Details_of_data_optimization} and \ref{sec:Hyperparameter_optimization_our_model}.
On the other hand, some objects do not capture the characteristics of Spitzer bubbles. For example, object id of 3, 4, 6, and 9 show strong \SI{24}{\micro m} and weak \SI{8}{\micro m} emission.
These false detections are likely due to the inclusion of the inappropriate Bubble data used for training. As explained in sub-subsection~\ref{sec:Processing_of_dataset}, the Bubble data is normalized to a range of 0 to 1 in areas other than point sources detected by DAOStarFinder. Subsequently, as described in subsection~\ref{sec:Effect_of_data_augmentation}, the Bubble data undergoes data augmentation, including enlargement. During this process, large bright objects may be included and, not being judged as point sources, are normalized as is. In such cases, the maximum value of the large bright object becomes 1, causing the emission from MWP-Bubbles to be underestimated. In particular, the \SI{8}{\micro m} emission from MWP-Bubbles is easily underestimated, resulting in the generation of Bubble data where \SI{24}{\micro m} emission is prominent. Due to this data, the model learns to identify point sources that are bright at \SI{24}{\micro m} as Spitzer bubbles, leading to the detection of point sources bright at \SI{24}{\micro m}, such as id = 3, 4, 6, and 9. Finding an appropriate method to remove objects that are compact at \SI{8}{\micro m} but brighter and more extended than point sources remains a challenge for future consideration. Id = 0 and 1 seems to be misidentified due to diffraction spikes at \SI{8}{\micro m} (\cite{2012SPIE.8442E..39H}).

Additionally, objects such as id = 101 and 106 were detected the same objects with different boxes. In SSD, BBoxes detected as bubbles with an overlap rate of 30\% or more are considered the same object. However, id = 101 and 106 have an overlap rate of less than 30\%. Some Spitzer bubbles are in contact with multiple Spitzer bubbles, making distinguishing them difficult based on simple numerical criteria.

\subsubsection{Application to training and validation region}
    \begin{table}[t]
    \caption
    {\label{table:MWP_match_rate} 
    Comparison of detection rates with ranked MWP-Bubble in the test region and the regions $1^{\circ} \leq |l| \leq 65^{\circ}$, $|b| \leq 1^{\circ}$.}
    \begin{center}
    \renewcommand{\arraystretch}{1.3}
    \begin{tabularx}{\columnwidth}{@{\extracolsep{\fill}}lccc}
    \toprule
    \multirow{2}{*}{}     & \multicolumn{3}{c}{\textbf{Rank}}  \\
                          & 1     & 1 + 2 & 1 + 2 + 3 \\ \midrule \midrule
    \textbf{Test Region}           & 98 \% & 78 \% & 60 \%     \\
    $\bm{1^{\circ} \leq |l| \leq 65^{\circ}, |b| \leq 1^{\circ}}$ & 97 \% & 81 \% & 63 \% \\ \bottomrule    
    \end{tabularx}
    \end{center}
    \vspace{-2pt}
    \end{table}
    \begin{table*}[t]
    \caption
    {\label{table:Test_catalogue} 
    A part of the catalog of the 1,413 objects that our model newly detected as Spitzer bubbles in the entire training and validation region. $G_{\text{LON}}$ and $G_{\text{LAT}}$ are the galactic longitude and latitude for the central position of the objects. Complete table~\ref{table:Test_catalogue} and the image of newly detected objects are available only on the online edition as the supplementary data in E-Table~\ref{table:Test_catalogue} and Figure E1.}
    \begin{center}
    \renewcommand{\arraystretch}{1.3}
    \begin{tabularx}{\textwidth}{@{\extracolsep{\fill}}crrr}
    \toprule
    \textbf{NAME}    & \textbf{G$_{\text{LON}}$ (deg)} & \textbf{G$_{\text{LAT}}$ (deg)} & \textbf{Radius (arcmin)} \\ \midrule \midrule
    $SB-GP_{0}$ & 5.7442 & 1.0089 & 0.22 \\
    $SB-GP_{1}$ & 14.3548 & 0.5842 & 0.22 \\
    $SB-GP_{2}$ & 295.5209 & -0.2372 & 0.23  \\ \bottomrule
    \end{tabularx}
    \end{center}
    \end{table*}
    \begin{figure*} [t]
    \begin{center}
    \begin{tabular}{c} 
    \includegraphics[width=16cm]{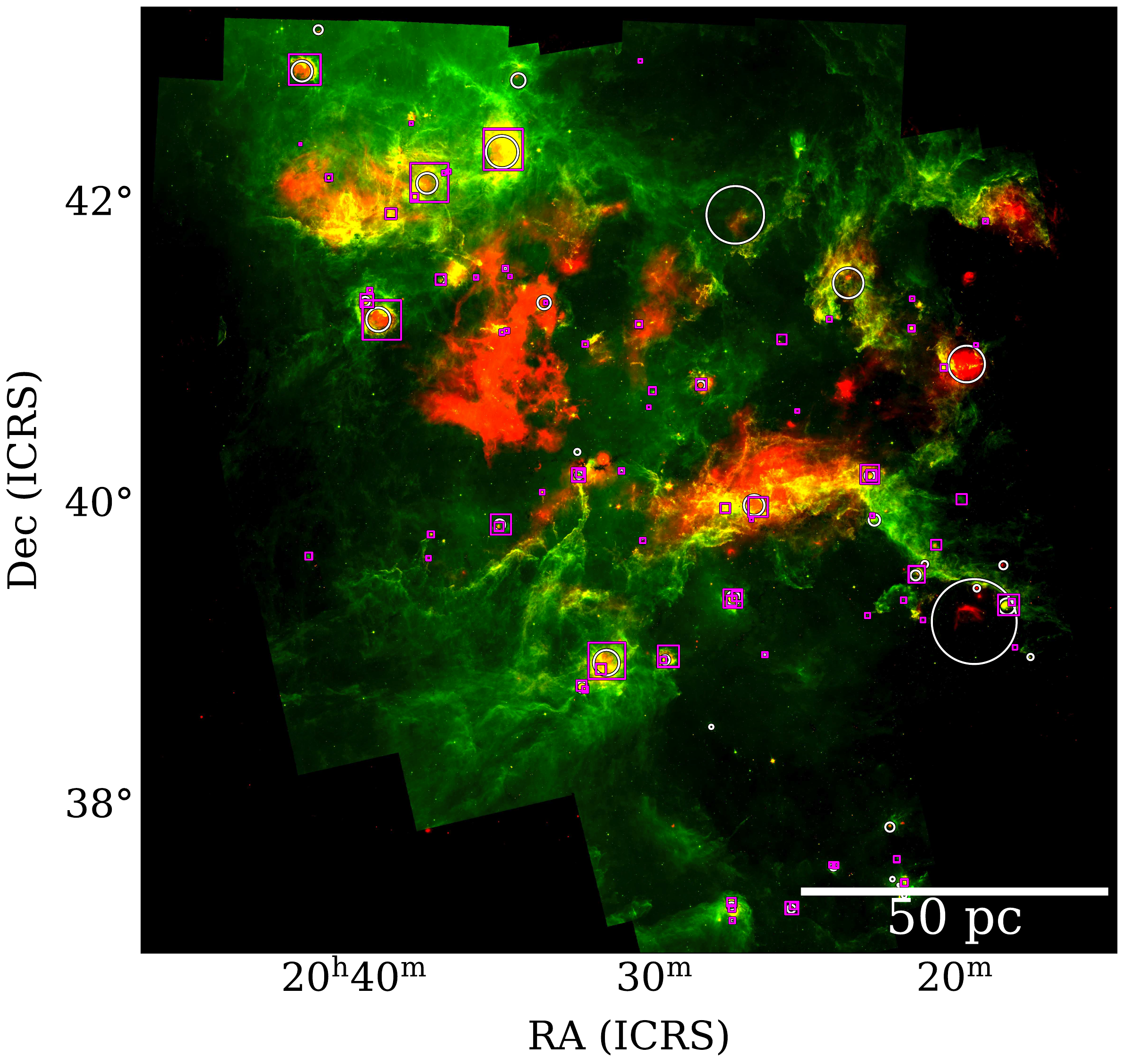}
    \end{tabular}
    \end{center}
    \caption
    {\label{fig:CygnusX_detection_map}
    The Cygnus $X$ with 69 objects detected by our model as Spitzer bubbles (magenta squares) and MWP-Bubble (white circles). The \SI{8}{\micro m} and \SI{24}{\micro m} emissions are shown in green and red, respectively.
    {Alt text: The composite Spitzer image of Cygnus $X$ overlaid with 69 objects detected by our model as Spitzer bubbles.
    }}
    \end{figure*}
    \begin{figure*} [t]
    \begin{center}
    \begin{tabular}{c} 
    \includegraphics[width=16cm]{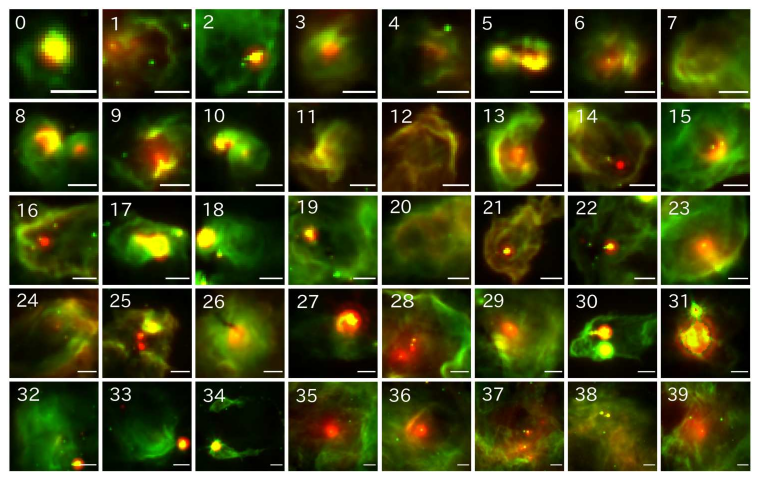}
    \end{tabular}
    \end{center}
    \caption
        { \label{fig:CygnusX_each_image} 
    The 40 objects newly detected as Spitzer bubbles by our model in the Cygnus $X$, sorted by size. The scale bar corresponds to 0.5 pc. The coordinate catalog of newly detected objects as Spitzer bubbles in the Cygnus $X$ are available only on the online edition as supplementary data in Table E1. The \SI{8}{\micro m} and \SI{24}{\micro m} emissions are shown in green and red, respectively. 
    {Alt text: Summary of images showing 40 newly detected objects in Cygnus $X$.}
    }
    \end{figure*}
We applied the model to the entire region, including the areas used for training and validation ($1^{\circ} \leq |l| \leq 65^{\circ}$, $|b| \leq 1^{\circ}$), to detect Spitzer bubbles that MWP could not detect. The model detected 3,006 objects, of which 1,413 were newly detected as bubbles. The detection rate for all MWP-Bubbles was almost the same as in the test region, and the detection rates for each rank were also almost identical (Table~\ref{table:MWP_match_rate}).

The detection rate for Rank 1 MWP-Bubbles was very high, while the detection rate decreased as Rank 2 and 3 MWP-Bubbles were included. The high detection rate for Rank 1 MWP-Bubbles and the low detection rate for Rank 2 and 3 MWP-Bubbles suggest that many newly detected sources have properties similar to Rank 1 bubbles. Indeed, many of the newly detected objects appear to be Spitzer bubbles. However, some newly detected objects are small and have point sources or fuzzy structures, indicating a certain number of false detections. With the current model performance, these need to be manually excluded. Table~\ref{table:Test_catalogue} shows part of the catalog of the newly detected bubbles.

The inference time for this region was approximately 3.6 hours, significantly shorter than MWP detection period. Given the high detection rate of Rank 1 MWP-Bubbles and the characteristics of the newly detected bubbles, it is evident that the deep learning model is highly effective in detecting Spitzer bubbles.

\subsubsection{Application to the Cygnus $X$ region}
We introduce the results of applying our model to Cygnus $X$. Cygnus $X$ is one of the most active star-forming regions in the Milky Way galaxy, located at a distance of approximately 1.4 kpc (\cite{2012A&A...539A..79R}). Cygnus $X$ contains hundreds of OB-type stars (\cite{2015MNRAS.449..741W}), and 47 MWP-Bubbles have been listed by \citet{2019MNRAS.488.1141J}. Schneider et al. (\yearcite{2006A&A...458..855S}, \yearcite{2007A&A...474..873S}) showed that the molecular clouds in Cygnus $X$ form connected groups, and it is understood that most of the molecular clouds in this region are at the same distance. We verified if it was possible to detect Spitzer bubbles in Cygnus $X$, which includes such massive molecular cloud complexes. The data used were survey data from IRAC and MIPS centered around (R.A.(J2000), Dec.(J2000)) = (20$^{h}$30$^{m}$25$^{s}$
, +\ang{40;00;}), covering an area of 24 deg$^2$ (\cite{2016ApJS..224...28P}).

Our model detected 69 objects in Cygnus $X$. Figure~\ref{fig:CygnusX_detection_map} shows the detected objects in magenta and MWP-Bubbles in white. It can be seen that objects of various sizes were detected. Table~\ref{table:confusion_matrix_Cygnus} shows the confusion matrix between the objects detected by our model and the MWP-Bubbles. The detection rate for all MWP-Bubbles was 62\%. Because the MWP-Bubbles in Cygnus $X$ were not ranked, the detection rate for all MWP-Bubbles was almost the same as that in the test region. Even in Cygnus $X$, where star formation is active, and the \SI{8}{\micro m} distribution is complex, the bubble detection rate of the model was close to that in the test region. 
    \begin{table}[t]
    \caption
    {\label{table:confusion_matrix_Cygnus} 
    Confusion matrix for Cygnus $X$. The MWP-Bubbles in Cygnus $X$ is 47, while our model detected 69 objects.
    }
    \begin{center}
    \renewcommand{\arraystretch}{1.3}
    \begin{tabularx}{\columnwidth}{@{\extracolsep{\fill}}|c|c|c|} \hline
    \diagbox{MWP}{Predicted} & Bubble & Non-Bubble \\ \hline
    Bubble & 29 & 18 \\ \hline
    Non-Bubble & 40 & - \\ \hline
    \end{tabularx}  
    \end{center}
    \end{table}
    \begin{figure*} [t]
    \begin{center}
    \begin{tabular}{c} 
    \includegraphics[width=16cm]{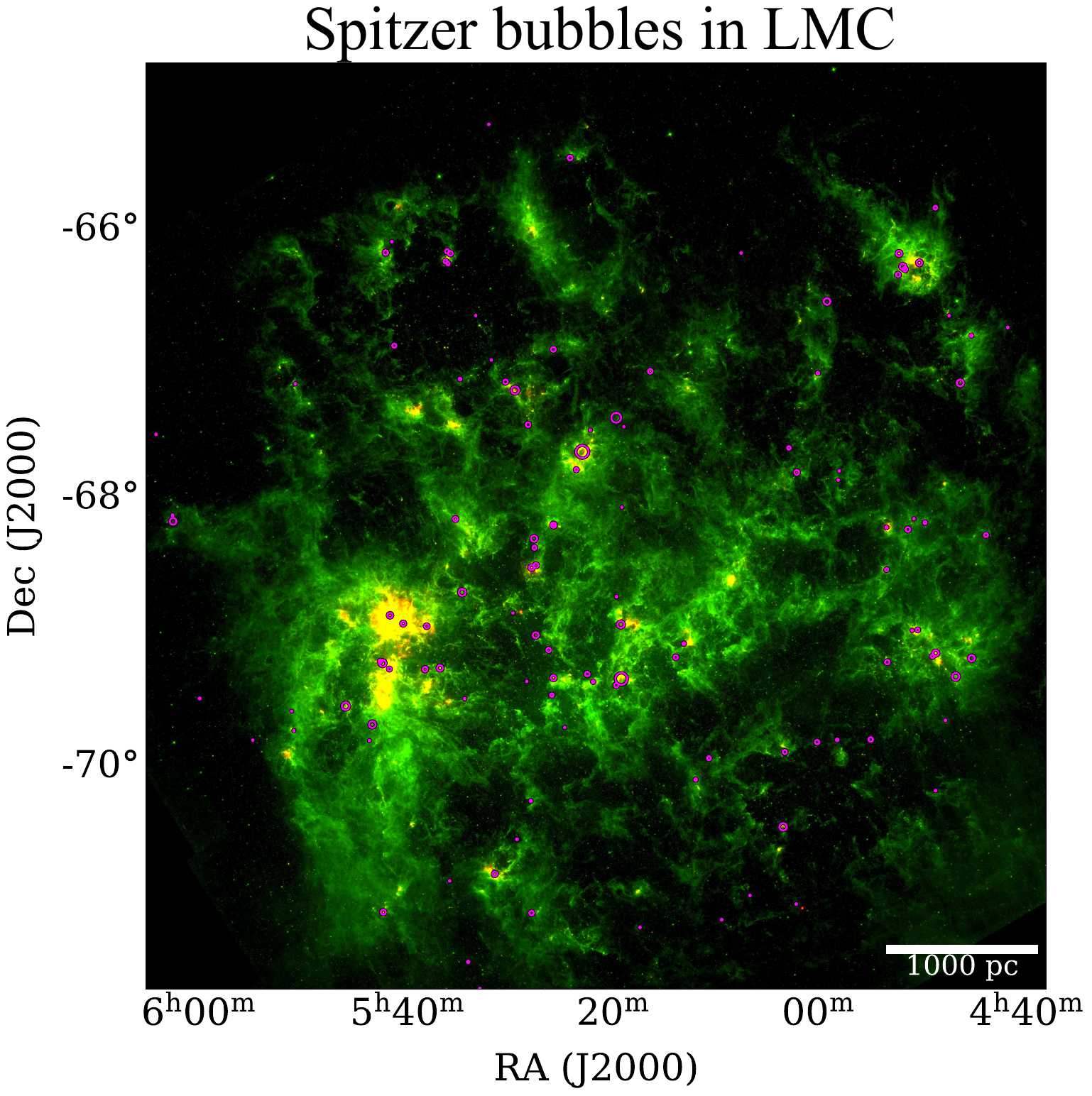}
    \end{tabular}
    \end{center}
    \caption
        { \label{fig:LMC_detection_map} 
    The LMC with 128 objects detected by our model as Spitzer bubbles (magenta circles). The \SI{8}{\micro m} and \SI{24}{\micro m} emissions are shown in green and red, respectively.
    {Alt text: The composite Spitzer image of the LMC overlaid with 128 objects detected by our model as Spitzer bubbles.
    }
    }
    \end{figure*}
    
Figure~\ref{fig:CygnusX_each_image} shows the 40 newly detected objects as bubbles (sorted by size). Many detected objects are bubbles where \SI{8}{\micro m} encloses \SI{24}{\micro m} (e.g., id = 3, 13, 21, 23, 26, 30, and 36). On the other hand, objects such as id = 7, 12, 20, and 38 are faint and extended in \SI{24}{\micro m}, similar to appearance at \SI{8}{\micro m}. These objects have shell-like structures, so the model judged them as bubbles, but it is difficult to determine whether they are Spitzer bubbles. 
Furthermore, objects such as id = 32 and 33, which have isolated \SI{8}{\micro m} distributions with strong \SI{24}{\micro m} point sources, were also detected. In such cases, it is necessary to improve the training data and increase the number of training iterations. The inference time for Cygnus $X$ was approximately 8 minutes.

\subsubsection{Application to LMC}
\label{sec:Application_to_LMC}
    \begin{figure*} [t]
    \begin{center}
    \begin{tabular}{c} 
    \includegraphics[width=16cm]{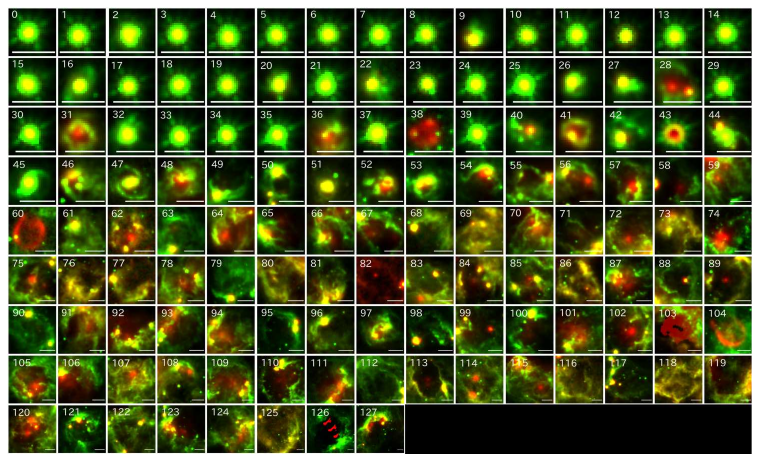}
    \end{tabular}
    \end{center}
    \caption
        { \label{fig:LMC_each_image} 
    The 128 objects newly detected as Spitzer bubbles by our model in the LMC, sorted by size. The scale bar is 10 pc. The coordinate catalog of newly detected objects as Spitzer bubbles in LMC is available only on the online edition as supplementary data in Table E2. The \SI{8}{\micro m} and \SI{24}{\micro m} emissions are shown in green and red, respectively.
    {Alt text: Summary of images showing 128 newly detected objects in the LMC.}
    }
    \end{figure*}

The Large Magellanic Cloud (LMC) is an extragalaxy located approximately 50 kpc from the Milky Way (\cite{2019Natur.567..200P}). Owing to small inclination (approximately 35$^\circ$: \cite{2001AJ....122.1807V}), individual objects in the LMC have less uncertainty in the distance. Additionally, the dust-to-gas mass ratio varies significantly spatially, being about 2--4 times the value near the Sun (\cite{2003ApJ...594..279G}). Observational studies in the LMC are important for investigating high-mass star formation in environments that are different from the Milky Way. However, the LMC is located farther away than objects in the Milky Way, resulting in worse spatial resolution compared to the test region.
No Spitzer bubble catalog has been made toward the LMC.
The data were taken from the Surveying the Agents of a Galaxy’s Evolution (\cite{2006AJ....132.2268M}), observed with IRAC (\SI{8}{\micro m}) and MIPS (\SI{24}{\micro m}).

Figure~\ref{fig:LMC_detection_map} shows the 128 objects identified as bubbles in magenta by the model. No object exceeding a few hundred pc was detected, but  a range of 10–-100 pc-sized objects were successfully identified.
Figure~\ref{fig:LMC_each_image} shows images of the 128 detected objects (sorted by size). It can be confirmed that many of the detected objects were \SI{24}{\micro m} enclosed by \SI{8}{\micro m}. Additionally, small objects such as id = 0--39 have characteristics similar to yellow balls (except for 28, 31, 36, and 38). The typical size of bubbles is less than several dozens of parsecs (see subsection~\ref{sec:Feature_of_Spitzer_bubble}), and considering the spatial resolution of the LMC (approximately 0.49 pc/pixel), these objects may be unresolved bubble structures.

However, it has been confirmed that some objects are associated with Mira-type variable stars (id = 6, 25, and 38) and T Tauri stars (id = 21). Objects associated with such stars have radiation spectra similar to Spitzer bubbles, making them difficult to distinguish. Furthermore, objects such as id = 69, 76, and 82, where the \SI{8}{\micro m} and \SI{24}{\micro m} distributions are similar to those seen in the Cygnus $X$ region, with an extended \SI{24}{\micro m} distribution resembling the appearance at \SI{8}{\micro m}, making it difficult to determine whether they were really Spitzer bubbles.

In addition to these characteristics, id = 44 (ESO 55-29), 45 (NGC 2150), and 47 (ESO 56-154) are galaxies or active galactic nuclei, and id = 60 and 104 (N132D) are SNRs. Galaxies have \SI{8}{\micro m} and \SI{24}{\micro m} emissions similar to Spitzer bubbles, making them difficult to distinguish; therefore, the model may have mistakenly detected them.

The inference time for the LMC was approximately 20 min.

\subsubsection{Application to NGC628}
\label{sec:Application_to_NGC628}
    \begin{figure*} [t]
    \begin{center}
    \begin{tabular}{c} 
    \includegraphics[width=16cm]{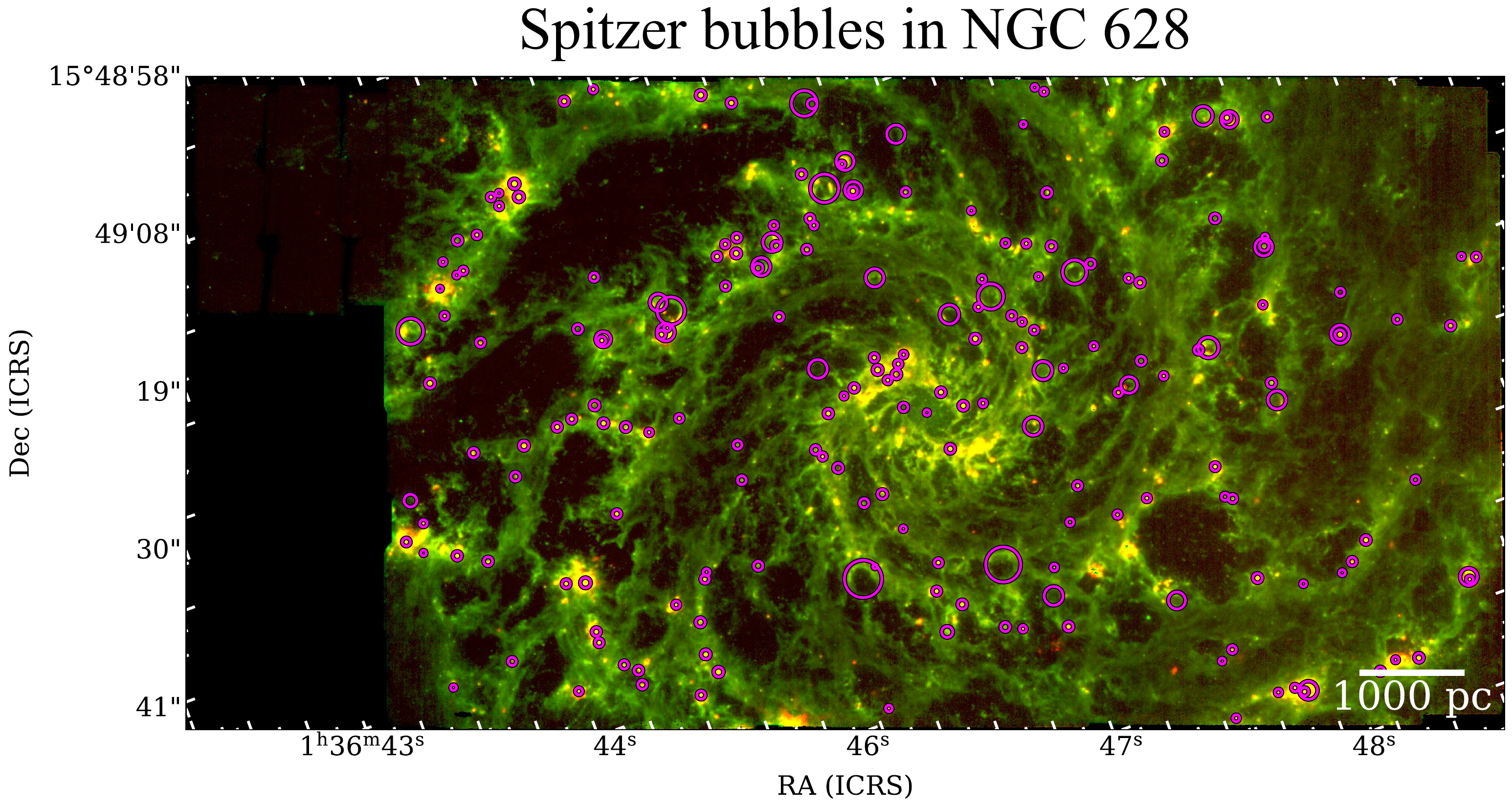}
    \end{tabular}
    \end{center}
    \caption
        { \label{fig:NGC628_detectoion_map} 
    JWST image of NGC 628 overlaid with 203 objects detected by our model as Spitzer bubbles (magenta circles). Many of the objects are distributed over the arm. The \SI{7.7}{\micro m} and \SI{21}{\micro m} emissions are shown in green and red, respectively.
    {Alt text: The composite JWST image with 203 objects detected by our model as Spitzer bubbles.}
    }
    \vspace{-14pt}
    \end{figure*}
    \begin{figure*} [t]
    \begin{center}
    \begin{tabular}{c} 
    \includegraphics[width=16cm]{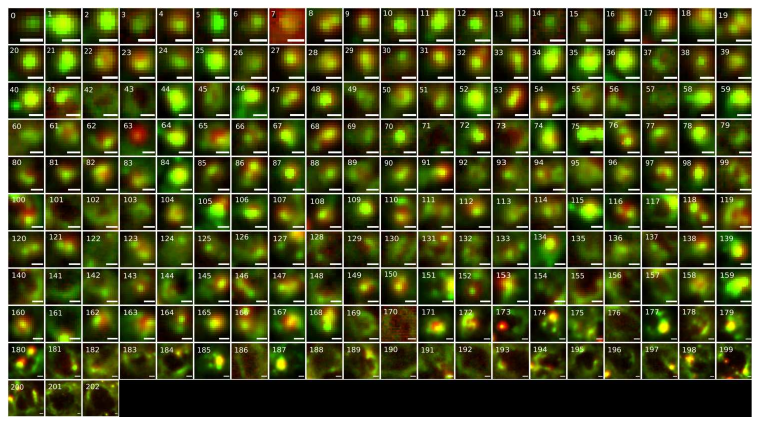}
    \end{tabular}
    \end{center}
    \caption
        { \label{fig:NGC628_each_image} 
    The 203 newly detected objects as Spitzer bubbles by our model in the NGC 628, sorted by size. Scale bar corresponds to 3 pc. The coordinate catalog of newly detected objects as Spitzer bubbles in the NGC 628 is available only on the online edition as supplementary data in Table E3. The \SI{7.7}{\micro m} and \SI{21}{\micro m} emissions are shown in green and red, respectively.
    {Alt text: Images of 203 newly detected objects in NGC 628.}
    }
    \end{figure*}
NGC 628 (also known as M74) is a spiral galaxy located 9.84 $\pm$ 0.63 Mpc from our galaxy. It is a face-on galaxy, allowing for detailed studies of spiral arms, star formation regions, and ISM. 
Since high-angular resolution observations have been made toward NGC 628 from optical to radio wavelengths, such as PHANGS-ALMA (Atacama Large Millimeter/submillimeter Array: \cite{2021ApJS..257...43L}), PHANGS-MUSE (Multi Unit Spectroscopic Explorer: \cite{2022A&A...659A.191E}), and PHANGS-HST (Hubble Space Telescope: \cite{2022ApJS..258...10L}), it is an ideal laboratory for understanding galaxy formation and evolution. PHANGS-JWST \citep{2023ApJ...944L..24W} detected 1964 Spitzer bubbles in NGC 628 using PHANGS data and JWST \SI{7.7}{\micro m}. We used JWST \SI{7.7}{\micro m} and \SI{21}{\micro m} data, which are close in wavelength to Spitzer \SI{8}{\micro m} and \SI{24}{\micro m}, to verify whether Spitzer bubbles can be detected in galaxies located farther than the LMC.

Since in distant galaxies such as NGC 628, the spatial resolution is more than 10 times worse than that in the LMC, many Spitzer bubbles with compact \SI{8}{\micro m} and \SI{24}{\micro m} distributions, such as yellow balls, may be detected. Therefore, for NGC 628, we added crop sizes of 25, 50, and 75 pixels to the inference crop sizes, resulting in 25, 50, 75, 100, 150, 300, 600, and 900 pixels. Crop sizes of 1200, 1500, 1800, 2400, and 3000 pixels were excluded as they would exceed the observation data. The sliding-window stride remained 1/3 of the crop size. One pixel of NGC 628 observation data from JWST is 0.11 arcsec, corresponding to 5.28 pc.

As a result of the inference, the model detected 203 objects as Spitzer bubbles. Figure~\ref{fig:NGC628_detectoion_map} shows the distribution of the detected objects. The sizes of the detected bubbles range from approximately 40 pc to 400 pc, with many bubbles distributed along the arms. It can be confirmed that many of the detected objects had \SI{24}{\micro m} enclosed by \SI{8}{\micro m} (Figure~\ref{fig:NGC628_each_image}). On the other hand, similar to the LMC, objects with compact \SI{8}{\micro m} and \SI{24}{\micro m} distributions were also observed owing to the low resolution.
The inference time for the NGC 628 was approximately 5 min.

\subsubsection{Characteristics of Spitzer bubbles detected by our model}
In this study, we ranked MWP-Bubbles and used only Rank 1 MWP-Bubbles for training and validation data. As shown in subsection~\ref{sec:Selection_of_Spitzer_bubbles}, using all MWP-Bubbles in the training data increases the number of obvious false detections that do not have the characteristics of Spitzer bubbles. 

The most concerning point about ranking MWP-Bubbles for use in training data is the potential omission of Spitzer bubbles generated by high-mass stars due to the restriction on the MWP-Bubbles. However, our developed model achieved a very high detection rate of 97\% for Rank 1 MWP-Bubbles. Additionally, the model could newly detect objects with similar characteristics to Rank 1 MWP-Bubbles.
The number of these newly detected objects is approximately equal to the number of Rank 1 MWP-Bubbles, effectively doubling the sample size. This fact demonstrates the limitations of human detection and highlights the importance of deep learning detection methods like those used in this study.

Furthermore, such comprehensive bubble detection is fundamental research for understanding star formation mechanisms, as seen in the statistical estimates of triggered star formation by \citet{2012MNRAS.421..408T} and \citet{2012ApJ...755...71K}.
    \begin{figure*} [t]
    \begin{center}
    \begin{tabular}{c} 
    \includegraphics[width=16cm]{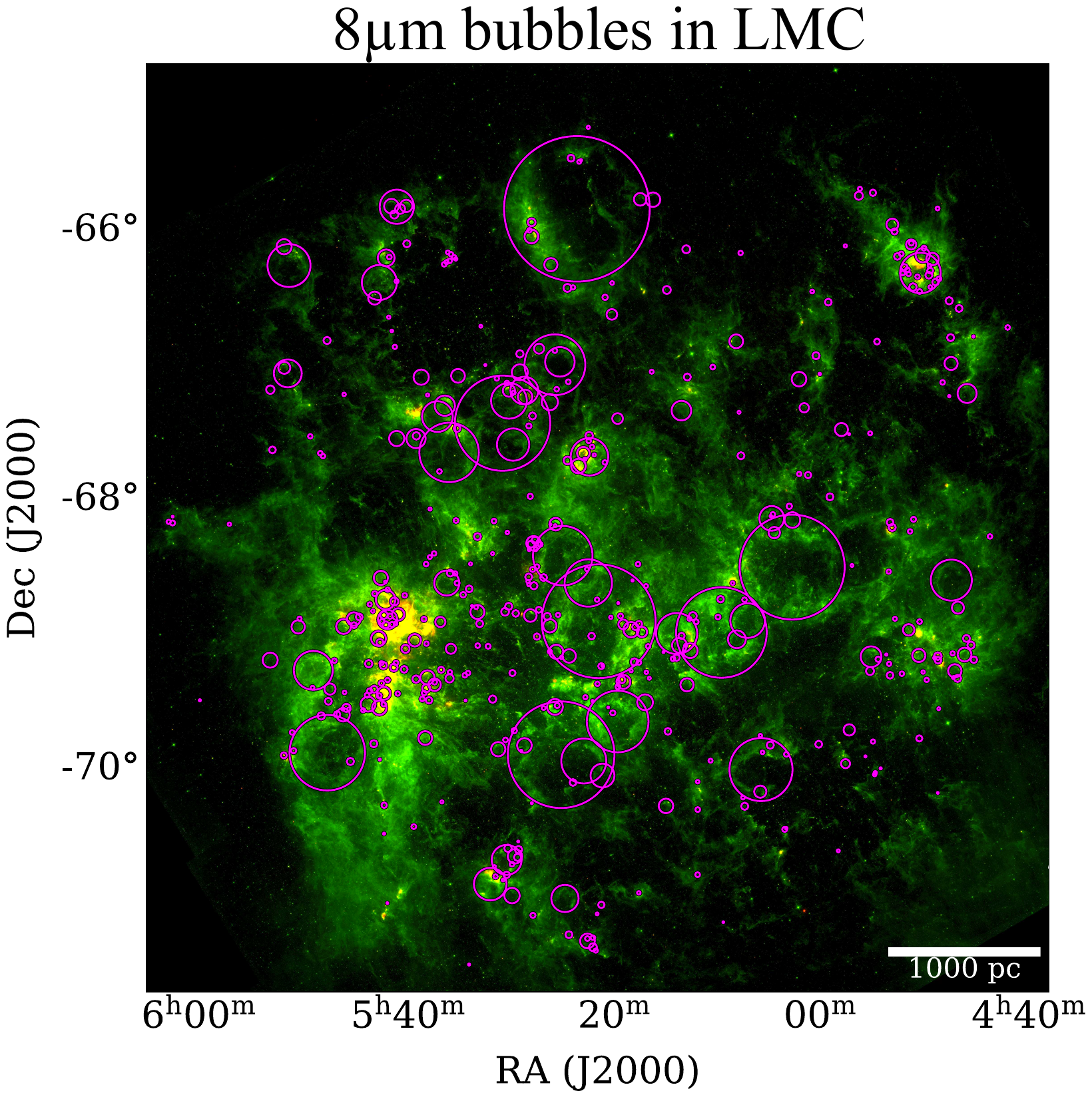}
    \end{tabular}
    \end{center}
    \caption
        { \label{fig:LMC_predict_8um} 
    The Spitzer image of the LMC overlaid with 469 objects detected by the model as \SI{8}{\micro m} shell-like structures (magenta circles). The image and coordinate catalog of the individual \SI{8}{\micro m} shell-like structures are available only on the online edition as supplementary data in Figure E2 and Table E4. The \SI{8}{\micro m} and \SI{24}{\micro m} emissions are shown in green and red, respectively.
    {Alt text: The composite Spitzer image of the LMC overlaid with 469 objects detected by the model as \SI{8}{\micro m} shell-like structures.}
    }
    \end{figure*}

\subsection{Detection of (super-) bubbles created by supernova}
In detecting Spitzer bubbles in the LMC and NGC 628, large shell-like structures observed in \SI{8}{\micro m} emission were rarely detected. This behavior reflects that these shell-like structures were not formed in association with recent high-mass star formation, as \SI{24}{\micro m} emission is not observed within the shells. In this subsection, we introduce the detection of shell-like structures observed in the \SI{8}{\micro m} emission band in the LMC and NGC 628 (hereafter, \SI{8}{\micro m} shell-like structures). Recent JWST observations have confirmed many \SI{8}{\micro m} shell-like structures in NGC 628 and other galaxies (\cite{2023ApJ...944L..22B, 2023MNRAS.521.5492M}). Shell-like structures larger than several hundred parsecs are thought to have been formed by supernova explosions. The ISM are swept by shock waves from supernova explosions approximately once every million years, and molecular clouds are born through repeated sweeping by shock waves. Filamentary molecular clouds perpendicular to the magnetic field are formed with each compression, and star formation begins when the line density becomes sufficiently large (\cite{2015A&A...580A..49I}). Many \SI{8}{\micro m} shell-like structures are considered to be formed through this process. We attempted to detect \SI{8}{\micro m} shell-like structures by applying the methods established in section~\ref{sec:Details_of_data_optimization} and \ref{sec:Hyperparameter_optimization_our_model}.
    \begin{figure*} [t]
    \begin{center}
    \begin{tabular}{c} 
    \includegraphics[width=16cm]{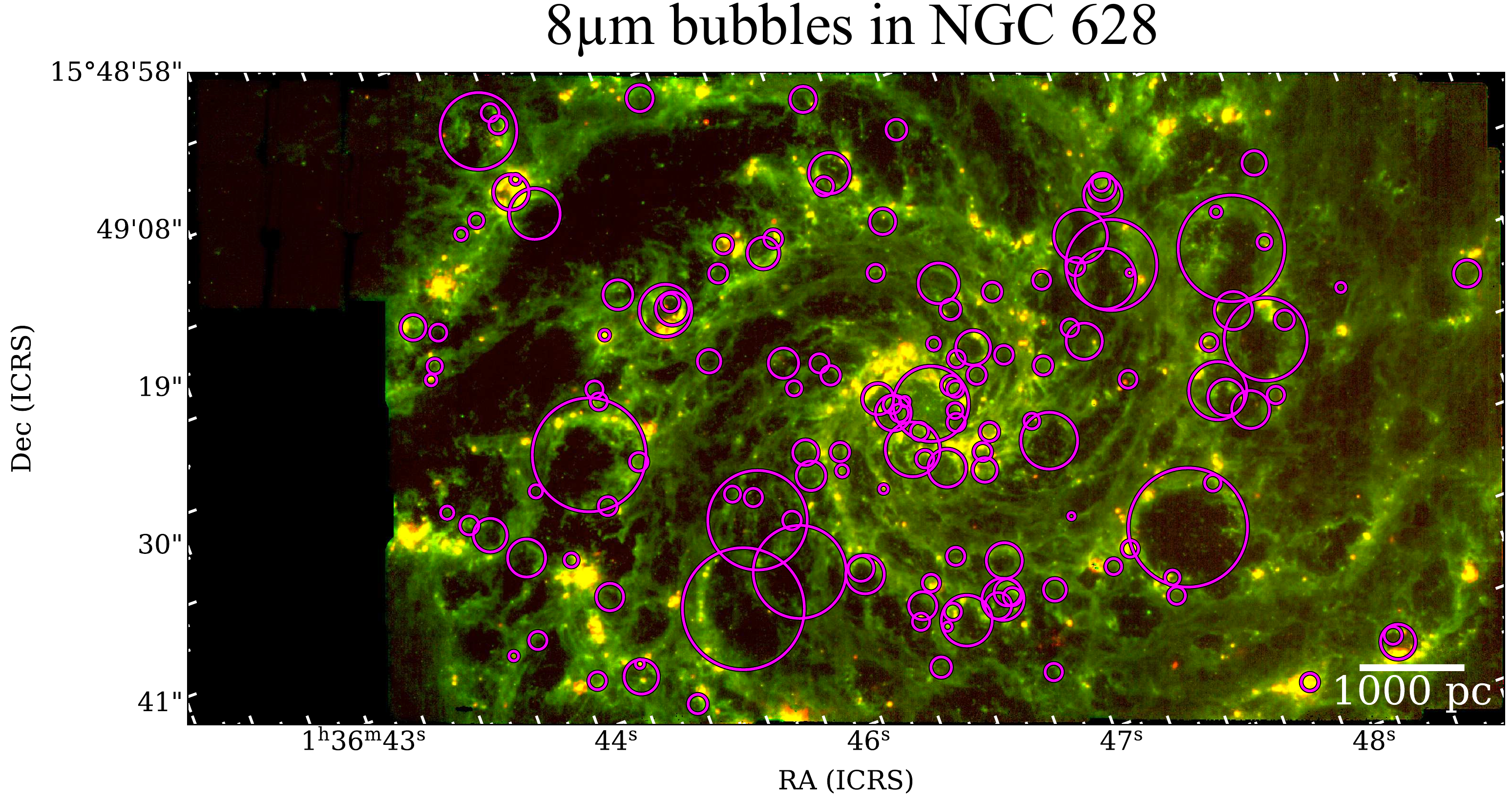}
    \end{tabular}
    \end{center}
    \caption
    { \label{fig:NGC628_8umShell} 
    The JWST image of NGC 628 overlaid with 143 objects detected by the model as \SI{8}{\micro m} shell-like structures (magenta circles). The image and coordinate catalog of the individual \SI{8}{\micro m} shell-like structures are available only on the online edition as supplementary data in Figure E3 and Table E5. The \SI{7.7}{\micro m} and \SI{21}{\micro m} emissions are shown in green and red, respectively.
    {Alt text: The composite JWST image of NGC 628 overlaid with 143 objects detected by the model as \SI{8}{\micro m} shell-like structures.}
    }
    \end{figure*}

We developed the model using the methods described in section~\ref{sec:Details_of_data_optimization} and \ref{sec:Hyperparameter_optimization_our_model}, changing only the training data. We created the training data by extracting only the \SI{8}{\micro m} data (Spitzer IRAC) from the training data used in section \ref{sec:Hyperparameter_optimization_our_model}. The MWP-Bubbles were reselected to include objects with well-confirmed \SI{8}{\micro m} shell-like structures (see subsection~\ref{sec:Selection_of_Spitzer_bubbles}). We note here that the training data we used are the \SI{8}{\micro m} shells associated with the Spitzer bubbles, not those created by the supernovae.

\subsubsection{Application to the LMC}
We trained a model capable of detecting \SI{8}{\micro m} shell-like structures and applied it to the LMC. As a result, the model detected 469 objects as \SI{8}{\micro m} shell-like structures (Figure~\ref{fig:LMC_predict_8um}). The sizes range from about 10 pc to about 900 pc. Most of these objects have well-confirmed \SI{8}{\micro m} shell-like structures. Some of these objects are accompanied by \SI{24}{\micro m} and are thought to be Spitzer bubbles. 

The crop size and sliding-window stride are the same as in sub-subsection~\ref{sec:Application_to_LMC}. The inference time was also about 20 min.

\subsubsection{Application to NGC 628}
Next, we introduce the results of applying this model to the \SI{7.7}{\micro m} data of NGC 628 observed by JWST. The crop size and the sliding-window stride were the same as those in sub-subsection~\ref{sec:Application_to_NGC628}. As a result, the model detected 143 objects as bubbles (Figure~\ref{fig:NGC628_8umShell}). Similar to the LMC, most objects had \SI{8}{\micro m} shell-like structures. Even with the model trained on the \SI{8}{\micro m} shell structure of Spitzer bubbles, which were formed by the emission of a young high-mass star, it is possible to detect \SI{8}{\micro m} shell-like structures on the order of a few hundred pc that may have been formed by a supernova explosion. However, some \SI{8}{\micro m} shell-like structures in NGC 628 were not comprehensively detected. This result may be caused because, while Spitzer bubbles typically exhibit a ring structure with an \SI{8}{\micro m} emission, some super-bubbles formed by supernova explosions may present just as holes in the extended ISM, which the current model may not detect. To detect bubble structures thought to be formed by supernova explosions, we need to construct training data by using data such as JWST data or simulation results, which is a subject of the forthcoming paper.

Recent observations indicate that galaxies are filled with bubbles formed by supernova explosions \citep{2023ApJ...944L..22B}. These bubbles interact with ISM and adjacent bubbles, inducing active star formation and forming Spitzer bubbles. Thus, by applying the method developed in this study to detect bubbles formed by supernova explosions observed at \SI{8}{\micro m}, it becomes possible to advance the study of star formation history in galaxies statistically by comparing the spatial distribution and the dynamics of the associated gas with these two types of bubbles.

\section{Summary}
We developed a deep learning model to detect Spitzer bubbles with the Single Shot Multibox Detector using 8 and \SI{24}{\micro m} data from the Spitzer Space Telescope. Applying this model to the Milky Way at $1^{\circ} \leq |l| \leq 65^{\circ}$, $|b| \leq 1^{\circ}$, we newly identified 1,413 objects as Spitzer bubbles, many of which exhibit distinct \SI{8}{\micro m} feature encompassing \SI{24}{\micro m} emission. In addition, the detection rate of Rank 1 MWP-Bubbles was very high at 98\%. When we also applied the model to Cygnus $X$, LMC, and NGC 628, the model newly detected 40 objects in Cygnus $X$, 128 in LMC, and 203 in NGC 628 as Spitzer bubbles. These newly detected objects shared similar characteristics to Rank 1 MWP-Bubbles, indicating that our model is effective in detecting Spitzer bubbles. Inference times varied by regions, requiring approximately 3.6 hours for the Milky Way ($1^{\circ} \leq |l| \leq 65^{\circ}$, $|b| \leq 1^{\circ}$), 8 min for Cygnus $X$, 20 min for LMC, and 5 min for NGC 628. Giving the high detection rate of Rank 1 MWP-Bubbles, the accuracy of newly detected bubbles, and the model's efficiency, this deep learning model approach proves effective for rapidly and accurately detecting Spitzer bubbles compared to manual method. We can use the model to detect Spitzer bubbles with high-speed and accuracy for observational data obtained by JWST and the other telescopes. However, some compact objects the model newly detected as bubbles are galaxies and Mira-type variable stars, which are difficult to distinguish from Spitzer bubbles. To further improve the performance of the model, we consider it necessary to create training data using simulations and construct an optimal architecture for detecting Spitzer bubbles in the future. 

The deep learning method used in this study can be applied to various objects. In this study, we attempted to detect shell-like structures formed in the \SI{8}{\micro m} band in LMC and NGC 628 using the Spitzer Space Telescope (\SI{8}{\micro m}) and JWST (\SI{7.7}{\micro m}) data. For the training data, we used the \SI{8}{\micro m} shell-like structure of Spitzer bubbles. The model detected 469 shell-like structures in LMC and 143 in NGC 628. Some of these \SI{8}{\micro m} shell-like structures may have been formed by supernova explosions. Although the model was able to detect many \SI{8}{\micro m} shell-like structures, there were still some objects that could not be detected. We consider it necessary to carefully create the training data for \SI{8}{\micro m} shell-like structures.

\begin{ack}
This work was supported by the 'Young interdisciplinary collaboration project' in the National Institutes of Natural Sciences (NINS) and JSPS KAKENHI Grant Numbers JP23H00129, JP18H05440. This work was also supported by JST SPRING, Grant Number JPMJSP2139. I would like to thank all the FUGIN-AI members who were involved in this study, and I would also like to thank Shota Ueda for his generous support. This work made use of Photutils, an Astropy package for detection and photometry of astronomical sources (\cite{larry_bradley_2023_7946442}).
\end{ack}

\section*{Supplementary data}
The following supplementary data is available in the online version of this article.

Table~\ref{table:Test_catalogue}, Tables E1--5, and Figures E1, E2, and E3.

\appendix
\section{Details of the machine and environment used in our model}
The machine and environment used for model development in this paper are as follows:

\subsection{Hardware Specifications}
\begin{description}
  \item [Machine Type :] Custom-built desktop
  \item [CPU :] Intel Core i9-13900K @ 5.50GHz, 24 cores 
  \item [GPU :] NVIDIA GeForce RTX 4090 @ 24GB GDDR6X
  \item [RAM :] 96GB DDR5 @ 4800MHz
  \item [Storage :] 2TB NVMe SSD
\end{description}

\subsection{Software Environment}
\begin{description}
  \item [Operating System :] Ubuntu 20.04.6 LTS
  \item [NVIDIA-Driver Version :] 545.23.08 
  \item [CUDA Version :] 12.3.107
  \item [Python Version :] 3.8.10
\end{description}

\subsection{Python Libraries}
\begin{description}
  \item [Astropy :] 5.2.2
  \item [Matplotlib :] 3.7.4
  \item [NumPy :] 1.24.4
  \item [Pandas :] 2.0.3
  \item [PyTorch :] 2.1.2
  \item [SciPy :] 1.10.1
\end{description}

The main Python Libraries are listed above. For other libraries that depend on these, see requirements.txt at the Github account.\footnotemark[3] Docker was also used to build the learning environment; see the docker folder at the Github account\footnotemark[3] for details on the Docker files.

\bibliographystyle{apj} 

\end{document}